\RequirePackage{etoolbox}
\csdef{input@path}{{style/}{graphics/}}
\documentclass[MSNbibl,nameyear,dvips]{arxstspdf}
\usepackage{dcolumn}
\usepackage{flushend}
\usepackage{stfloats}

\volume{30}
\issue{1}
\pubyear{2015}
\firstpage{96}
\lastpage{117}
\doi{10.1214/14-STS503}% Updated by VTEXPTS2LaTeX.exe, 09.12.2014 10:15
\docsubty{FLA}

\makeatletter

\newcolumntype{d}[1]{D{.}{.}{#1}}
\newcommand{\lleft}{\left}
\newcommand{\rright}{\right}
\newcommand{\logit}{\operatorname{logit}}
\newcommand{\expit}{\operatorname{expit}}
\newcommand{\diag}{\operatorname{diag}}
\makeatother

\begin{document}
\begin{frontmatter}

\title{Estimating Structural Mean Models with Multiple Instrumental
Variables Using the Generalised Method of Moments}%\thanksref{T1}
% kai straipsnis turi susijusiu diskusiju ir rejoinder'iu
%rejoinder at \relateddoi{r}{10.1214/00-STSXXXX}.}
\runtitle{Estimating SMM using GMM}

\begin{aug}
% Corresponding author: Paul Clarke - pclarke@essex.ac.uk% Updated by
%VTEXPTS2LaTeX.exe, 09.12.2014 10:15
\author[A]{\fnms{Paul S.}~\snm{Clarke}\corref{}\ead[label=e1]{pclarke@essex.ac.uk}},
\author[B]{\fnms{Tom M.}~\snm{Palmer}\ead[label=e2]{T.M.Palmer@warwick.ac.uk}}
\and
\author[C]{\fnms{Frank}~\snm{Windmeijer}\ead[label=e3]{f.windmeijer@bristol.ac.uk}}
\runauthor{Clarke, Palmer and Windmeijer}

\affiliation{University of Essex, University of Warwick and University
of Bristol}

\address[A]{Paul S. Clarke is Professor, Institute for Social and
Economic Research, University of
Essex, Wivenhoe Park, Colchester CO4 3SQ, UK \printead{e1}.}
\address[B]{Tom M. Palmer is Assistant Professor, Division of Health
Sciences, Warwick
Medical School, University of Warwick, Coventry CV4 7AL, UK \printead{e2}.}
\address[C]{Frank Windmeijer is Professor, Centre for Market and Public
Organisation  and~Department of Economics, University of Bristol, 8~Woodland Road, Bristol BS8 1TN,
UK \printead{e3}.}
\end{aug}

% ABSTRACT
%
\begin{abstract}
Instrumental variables analysis using genetic markers as instruments is now
a widely used technique in epidemiology and biostatistics. As single markers
tend to explain only a small proportion of phenotypic variation, there is
increasing interest in using multiple genetic markers to obtain more precise
estimates of causal parameters. Structural mean models (SMMs) are
semiparametric models that use instrumental variables to identify causal
parameters. Recently, interest has started to focus on using these models
with multiple instruments, particularly for multiplicative and logistic
SMMs. In this paper we show how additive, multiplicative and logistic SMMs
with multiple orthogonal binary instrumental variables can be estimated
efficiently in models with no further (continuous) covariates, using the
generalised method of moments (GMM) estimator. We discuss how the Hansen
\textit{J}-test can be used to test for model misspecification, and how
standard GMM software routines can be used to fit SMMs. We further show that
multiplicative SMMs, like the additive SMM, identify a weighted average of
local causal effects if selection is monotonic. We use these methods to
reanalyse a study of the relationship between adiposity and hypertension
using SMMs with two genetic markers as instruments for adiposity. We find
strong effects of adiposity on hypertension.
\end{abstract}

% KEYWORDS
\end{frontmatter}
%

%s1 #&#
\section{Introduction}

Additive and multiplicative structural mean models (SMMs) and G-estimation
were introduced by Robins (\citeyear{Rob89}, \citeyear{Rob94}) for estimating the causal
effects of
treatment regimes on outcomes from encouragement designs, namely, randomised
controlled trials (RCTs) affected by noncompliance. Additive SMMs are
parameterised in terms of average treatment effects and multiplicative SMMs
in terms of causal risk ratios; the G-estimators for these models are
consistent, asymptotically normal and can be constructed to be
semiparametrically efficient. \citet{VanGoe03}
subsequently developed a class of estimators for generalised SMMs and, in
particular, the ``double-logistic'' SMM for estimating causal odds ratios.
Within this literature, causal effects among the treated are identified by
the assumption of no effect modification by the instrumental variable (NEM),
that is, the causal effect among the treated is the same at each level of
the instrumental variable; see, for example, \citet{HerRob06}.
Alternative estimators and identifying assumptions for generalised SMMs have
also been developed by \citet{RobRot04}, \citet{Tan10} and, for a
closely related class of models, \citet{vanHubJew07}.

The application of SMMs is not limited to encouragement designs, however,
and extends to the analysis of observational studies using instrumental
variables; see, for example, \citet{HerRob06}. Instrumental
variables
analysis involves estimating the causal effect of a temporally antecedent
predictor variable on an outcome using an instrumental variable that is
associated with the outcome \textit{only} through its association with the
predictor. Instrumental variables analysis has historically been a
domain of
econometrics, but is now frequently used within epidemiology and
biostatistics. In particular, genetic markers were proposed as instruments
for modifiable risk factors by \citet{Kat86} and \citet{DavEbr03}. Epidemiological studies using genetic markers are known as Mendelian
randomisation studies after the assumption that each individual's genotype
is randomly assigned at conception, which implies that the genetic
marker is
an instrumental variable if it at least partly explains variation in the
risk factor. In practice, genetic markers explain only a small
proportion of
phenotypic variation, and so large sample sizes are required to obtain any
reasonable precision. The number of genome-wide association studies has
increased as the costs of genotyping have decreased, which has led to the
identification of multiple genetic variants for the same risk factor. An
important attraction of using multiple genetic variants as instrumental
variables is that, potentially, more precise causal estimates can be
obtained.

Techniques for multiple instruments in linear instrumental variables
analysis are already in use; see, for example, \citet{Paletal12}.
For linear and
nonlinear SMMs, the different frameworks we have mentioned are all general
enough to incorporate multiple instrumental variables, but to date the focus
in applications has mainly been on cases involving a single instrumental
variable. The exceptions are \citet{BowVan11} and \citet{Tan10}.
In the first paper, within the frameworks introduced by \citet{Rob94} and
\citet{VanGoe03}, the authors propose a combination of
multiple instrumental variables into a single instrumental variable which,
they argue, leads to an optimally efficient estimator. In the second paper,
multiple instrumental variables are directly incorporated into the
estimating equations, within an alternative framework that introduces new
structural models together with doubly robust estimating equations.

In this paper, we consider an alternative framework based on the generalized
method of moments (GMM); see, for example, \citet{Han82} and \citet{New93}. GMM is
widely used in econometrics for the estimation of instrumental variables
models. We show how nonlinear SMMs with multiple instruments can be
formulated as instrumental variables models and estimated using GMM.
Furthermore, if the instrumental variables result in an over-identified
model, then the Hansen \textit{J}-test can be used to test parametric
identifying assumptions like NEM. We also argue that GMM has good efficiency
properties for SMMs without baseline covariates. Specifically, GMM is shown
to be semiparametrically efficient in cases where the instrumental
variables can be represented by a set of orthogonal binary variables, in
which case the efficient combination of the instrumental variables is
equivalent to that proposed by \citet{BowVan11}. An important
practical advantage of GMM is that it can be implemented using existing
routines in software packages like \textit{Stata} and \textit{R}; see
Chauss\'{e} (\citeyear{Cha10}).

The focus of our presentation is on SMMs without covariates because these
models are widely applicable to Mendelian randomisation studies. A drawback
to fitting SMMs with covariates using our approach is that the user must
correctly specify the covariate effects in a model for the counterfactual
exposure-free outcomes, which cannot be tested for misspecification.
However, if the covariate effects are saturated---in the sense that the
covariates define population strata and the SMM has a separate parameter
for the causal effect in each stratum---then this counterfactual model is
nonparametric and cannot be misspecified, and the efficiency properties
listed above all hold. Saturated SMMs like this can be used to deal
with population stratification in Mendelian randomisation studies; see,
for example, \citet{Lawetal08}. \citet{Tan10} also uses GMM but applies
it to a
very different family of doubly robust estimating equations for which the
user must specify the covariate effects in two sets of models; the advantage
of this approach is that each model can be tested for misspecification, and
the estimator remains consistent for the SMM parameters even if one set of
models is misspecified.

In the second part of the paper, we consider the interpretation of additive
and multiplicative SMMs with multiple instruments when the key NEM
assumption fails. In such circumstances, an additive SMM with one binary
instrument identifies a ``local'' average treatment effect (LATE)---also known
as a ``complier'' average causal effect (CACE)---provided that selection is
monotonic, and multiplicative SMMs identify local causal risk ratios;
see, for example, \citet{ClaWin10}. When there are multiple
instruments,
\citet{ImbAng94} show that a GMM estimator for the additive
SMM\
identifies a weighted average of LATEs. We extend their analysis to
multiplicative SMMs to show that a GMM estimator identifies weighted
averages of local risk ratios.

To demonstrate our findings, we reanalyse data from a study of the
relationship between hypertension and adiposity by \citet{Timetal09}. In
the original study, two genetic markers were used as instruments for
adiposity and analysed using linear instrumental variables models. We
reanalyse this study by focusing on hypertension as a binary outcome and
by estimating causal effects of adiposity using multiplicative and logistic
SMMs.

The remainder of the paper is organised as follows. In Section~\ref{sec2} we review
the potential outcomes framework and the additive, multiplicative and
logistic SMMs, first for the simple case of a single binary instrumental
variable and then more generally. In Section~\ref{sec3} we show how SMMs with a
single binary instrument can be formulated as an instrumental variables
model and estimated using GMM, and in Section~\ref{sec4} extend this to multiple
instrumental variables. In Section~\ref{sec5} we discuss how GMM combines multiple
instruments efficiently for orthogonal binary instruments. In Section~\ref{sec6} we
present the results of a Monte Carlo study for multiplicative and logistic
SMMs. In Section~\ref{sec7} we derive the multiple instruments results for the local
risk ratio. Finally, in Section~\ref{sec8} we apply our estimation procedures to
reanalyse the adiposity and hypertension data of \citet{Timetal09}, and
in Section~\ref{sec9} make concluding remarks. In the \hyperref[app]{Appendix} we provide Stata and
R code for the estimation of the three SMMs using GMM.

%s2 #&#
\section{Structural Mean Models}\label{sec2}

%s2.1 #&#
\subsection{The Basic Setup}\label{sec2.1}

To introduce SMMs, we follow the exposition in \citet{HerRob06}
and focus on SMMs for a randomised controlled trial where $Z_{i}$, $X_{i}$
and $Y_{i}$ are i.i.d. dichotomous random variables for individual
subjects $i=1,\ldots,n$ drawn from the target population. For
individual $i$, let $Z_{i}$ be a binary indicator of treatment assignment following
randomization, $X_{i}$ the selected treatment, and $Y_{i}$ the study
outcome. For notational simplicity the subject index is sometimes suppressed
for the random variables.

The potential outcomes can now be defined in the usual way. The potential
treatments $X_{0}$ and $X_{1}$ are the treatments selected by the individual
following assignment to treatment $z=0,1$, respectively. Similarly, the
potential (study) outcome $Y_{xz}$ is that obtained if the individual is
assigned to treatment $z$ but given treatment~$x$. Using potential outcomes
notation, we can now state five key conditions that must be satisfied for
causal inference: (i) the ``stable unit treatment value assumption'' that
each individual's potential treatments and potential study outcomes are
mutually independent of those for any other individual; (ii) the
``consistency assumption'' $X=X_{Z}$ and $Y=Y_{XZ}$ that links the observed
realisations to the potential outcomes; (iii) the ``independence assumption'',
potential outcomes $Y_{zx}$ are independent of $Z$; (iv) the ``exclusion
restriction'' $Y_{xz}=Y_{x}$; and (v) ``association assumption'', there is an
association between $X$ and $Z$. Alternative statements of the key
conditions can be found in \citet{RobRot04} and \citet{Tan10}.

%s2.2 #&#
\subsection{SMM Identification}\label{sec2.2}

For the basic setup defined above, the generalized SMM of \citet{VanGoe03} is
%
%e1 #&#
\begin{eqnarray}\label{GSMM}
&&h\bigl\{E(Y|X,Z)\bigr\}-h\bigl\{E(Y_{0}|X,Z)\bigr\}\nonumber\\[-8pt]\\[-8pt]
&&\quad = (
\psi_{0}+\psi_{1}Z ) X,\nonumber
\end{eqnarray}
where $Y_{0}$ is often referred to as the exposure-free potential outcome,
and $h$ is the link function that determines the interpretation of the
target causal parameters $\psi_{0}$ and $\psi_{0}+\psi_{1}$. For example,
the identity link leads to the additive SMM
$E(Y|X,Z)-E(Y_{0}|X,Z)= (
\psi_{0}+\psi_{1}Z ) X$, where $\psi
_{0}=E(Y_{1}-Y_{0}|X=1,Z=0)$ and
$\psi_{0}+\psi_{1}=E(Y_{1}-Y_{0}|X=Z=1)$ are both average treatment
effects; the log link leads to the multiplicative SMM $%
E(Y|X,Z)/E(Y_{0}|X,Z)=\exp\{ ( \psi_{0}+\psi_{1}Z ) X\}$,
where $%
\exp(\psi_{0})=E(Y_{1}|X=1,Z=0)/\allowbreak E(Y_{0}|X=1,Z=0)$ and $\exp(\psi
_{0}+\psi_{1})=E(Y_{1}|X=Z=1)/E(Y_{0}|X=Z=1)$ are causal risk ratios.

The SMM parameters are identified by exploiting the conditional mean
independence (CMI), or randomisation, assumption
%
%e2 #&#
\begin{equation}
E(Y_{0}|Z)=E(Y_{0}), \label{cmi}
\end{equation}
which follows automatically from the key conditions on $Z$ specified above.
For the additive SMM, $h$ is the identity link and $E(Y_{0}|Z)=E\{
Y- (
\psi_{0}+\psi_{1}Z ) X|Z\}$; and for the multiplicative SMM,
$h=\log$
and $E(Y_{0}|Z)=E [ Y\exp \{ -(\psi_{0}+\psi_{1}Z)X
\} |Z%
 ] $. However, the CMI assumption (\ref{cmi}) alone does not identify
$\psi_{0}$ and $\psi_{1}$; for instance, in this simple setup, CMI
implies the single independent moment condition
%
%e3 #&#
\begin{eqnarray}\label{MCAdd}
&&E\bigl\{Y- ( \psi_{0}+\psi_{1} ) X|Z=1\bigr\}\nonumber\\[-8pt]\\[-8pt]
&&\quad =E(Y-
\psi_{0}X|Z=0),\nonumber
\end{eqnarray}
under the additive SMM. In other words, there is one moment condition with
two unknowns. Hence, we must impose dimension-reducing constraints on the
SMM. \citet{HerRob06} highlight the importance of no effect
modification by $Z$ (NEM), which constrains $\psi_{1}=0$ in (\ref
{MCAdd})
and identifies $\psi_{0}$. Under NEM, the parameter $\psi_{0}$ of the
additive SMM can be interpreted as $E(Y_{1}-Y_{0}|X=1)$, that is, the
average causal effect among the treated; and the parameter $\exp(\psi_{0})$
of the multiplicative SMM can be interpreted as
$E(Y_{1}|X=1)/E(Y_{0}|X=1)$, that is, the causal risk ratio among the treated.

Generally, the form of $E(Y_{0}|Z)$ is more complex than for the additive
and multiplicative SMMs because the inverse link function $h^{-1}$ is not
separable. Specifically, for the additive SMM, $h=h^{-1}$ is the additively
separable identity function [i.e., $h^{-1}(a+b)=h^{-1}(a)+h^{-1}(b)$]; and
for the multiplicative SMM, $h=\log$ so that $h^{-1}=\exp$ is
multiplicatively separable [i.e., $h^{-1}(a+b)=h^{-1}(a)\times h^{-1}(b)$].
For nonseparable~$h^{-1}$, however, CMI and NEM do not alone identify the
parameters of SMMs. For example, the logistic SMM
%
%e4 #&#
\begin{eqnarray}\label{logisticSMM}
&&\logit \bigl\{E(Y|X,Z)\bigr\}-\logit \bigl\{E(Y_{0}|X,Z)
\bigr\}\nonumber\\[-8pt]\\[-8pt]
&&\quad = ( \psi _{0}+\psi _{1}Z ) X,\nonumber
\end{eqnarray}
where $\logit (p)=\log\{p/(1-p)\}$ and the parameters $\exp
(\psi
_{0}) $ and $\exp(\psi_{0}+\psi_{1})$ are causal odds ratios for the
$%
(X,Z)=(1,0)$ and $(1,1)$ groups, respectively; assuming that CMI and NEM
hold,
%
%e5 #&#
\begin{eqnarray} \label{MCLogNEM}
E(Y_{0}|Z)&=&E\bigl[\expit \bigl\{\logit  \bigl( E(Y|X,Z)
\bigr)\nonumber\\[-8pt]\\[-8pt]
&&\hspace{81pt}{} -\psi _{0}X\bigr\}|Z\bigr],\nonumber
\end{eqnarray}
where $\expit (a)=\exp(a)/\{1+\exp(a)\}$ is the nonseparable inverse
logit function. It is clear that $\psi_{0}$ is not identified unless $%
E(Y|X,Z)$ is known; see, for example, \citet{Rob00}. Hence, to identify
$\psi_{0}$,
it is necessary to specify an association model
%
%e6 #&#
\begin{equation}
h_{a} \bigl\{ E(Y|X,Z) \bigr\} =m_{\bolds{\beta}}(X,Z), \label{AssocMod}
\end{equation}
where $h_{a}$ is its link function and $m_{\bolds{\beta}}(X,Z)$ its linear
predictor. \citet{VanGoe03} specify the double-logistic
SMM such that $h_{a}=h=\logit $, where the SMM parameters are
identified by the conditional moment conditions
\begin{eqnarray*}
&&E\bigl[\expit \bigl\{m_{\bolds{\beta}}(X,Z)-\psi_{0}X\bigr\}|Z=0
\bigr]\\
&&\quad =E\bigl[\expit %
\bigl\{m_{\bolds{\beta}}(X,Z)-
\psi_{0}X\bigr\}|Z=1\bigr],
\\
&& E \bigl[ Y-\expit  \bigl\{ m_{\bolds{\beta}}(X,Z) \bigr\} |X,Z \bigr] =0,
\end{eqnarray*}
provided that the association model is correctly specified. A saturated
association model is $m_{\bolds{\beta}}(X,\break Z)=\beta_{0}+\beta
_{1}X+\beta
_{2}Z+\beta_{3}XZ$ for the simple setup considered here, and is
nonparametric in the sense of placing no constraints on the
distribution of
the observed data.
However, nonsaturated logistic association models are potentially uncongenial to the
logistic SMM and hence misspecified; see \citet{RobRot04}.
\citet{RobRot04} propose an
estimator that solves this problem, but \citet{Vanetal11} argue
that the impact of an uncongenial association model will be small in
practice.

As highlighted by \citet{VanGoe05} and \citet{Tan10}, for
more general scenarios where any or all of $X$, $Z$ and $Y$ are nonbinary,
NEM is not the only identifying assumption for SMMs. For example, if
$Z$ has
three categories and $X$ is binary, then CMI implies $3$ independent moment
conditions, and so the model can be identified if it is correct to assume
that $Z$ has a linear effect and the SMM is $ ( \psi_{0}+\psi
_{1}Z ) X$, which identifies both SMM parameters without needing to
assume NEM.

%s2.3 #&#
\subsection{Estimating Equations}\label{sec2.3}

The construction of consistent estimating equations requires the
specification of suitable \textit{unconditional} moment conditions
based on
the \textit{conditional }moment conditions introduced above. The estimating
equations are sample analogues of these unconditional moment
conditions, and
the different estimating approaches in the SMM literature differ in how
these unconditional moment conditions are specified. We first consider
estimating equations for simple scenarios involving only binary variables,
before moving on to the more general case.

\citet{Rob94} derived G-estimation for additive and multiplicative
SMMs. The
G-estimator is based on an unconditional moment condition of the form
%
%e7 #&#
\begin{equation}
E \bigl[ \bigl\{ Z-E(Z) \bigr\} E(Y_{0}|Z) \bigr] =0, \label{GestMC}
\end{equation}
which holds under (\ref{cmi}). As shown above, for SMMs with separable
inverse link functions, we can write $E(Y_{0}|Z)=E\{h^{\ast}(X,Y;\psi
_{0})|Z\}$, where $h^{\ast}$ is determined by the SMM and NEM is taken to
hold. Thus, the sample analogue of (\ref{GestMC}) is
%
%e8 #&#
\begin{equation}
n^{-1}\sum_{i=1}^{n} \bigl\{ Z_{i}-E(Z)
\bigr\} h^{\ast
}(X_{i},Y_{i};\widehat{
\psi}_{0})=0, \label{GEstEq}
\end{equation}
where, for example, $h^{\ast}(X,Y;\widehat{\psi}_{0})=Y-\widehat
{\psi}%
_{0}X$ for the additive SMM and $h^{\ast}(X,Y;\widehat{\psi
}_{0})=Y\exp(-%
\widehat{\psi}_{0}X)$ under the multiplicative SMM. Under regularity
conditions, $\widehat{\psi}_{0}$ is a consistent estimator for $\psi_{0}$
under CMI provided that (a) the SMM is correctly specified and (b)~$E(Z)$ is
known. The second of these conditions will be satisfied if $Z$ is based
on a
known allocation rule such as randomisation. Otherwise, if $E(Z)$ is
unknown, we must specify a (trivial) model $E(Z)=\mu$ and replace
$E(Z)$ in
(\ref{GEstEq}) with $\widehat{\mu}$, that is, a consistent estimator
of $\mu$. \citet{RobMarNew92} note that the correct asymptotic covariance
matrix for $\widehat{\psi}_{0}$ can only be derived from an extended system
of moment conditions that includes $E(Z-\mu)=0$; see also \citet{VanGoe03} and \citet{Tan10}. Conversely, treating $\widehat{\mu
}$ as
known when deriving the asymptotic variance of $\widehat{\psi}_{0}$ leads
to an expression that is too large, and results in conservative inferences;
see \citet{RobMarNew92} and \citet{VanGoe03}.

The estimating equations for the double-logistic SMM are
%
%e9 #&#
\begin{equation}
\quad n^{-1}\sum_{i=1}^{n}(Z_{i}-\mu)
\expit \bigl\{m_{\bolds{\beta}%
}(X,Z)-\psi_{0}X\bigr\}=0,
\label{DLSMMEstEq}
\end{equation}
where $\mu=E(Z)$ as before. Due to the nonseparability of the $\expit
$ function, the estimating equation involves the association model
$\logit\{E(Y|X,Z)\}=m_{\bolds{\beta}}(X,Z)$. As with $\mu$, we must
replace $\bolds{\beta}$ in (\ref{DLSMMEstEq}) with a consistent
estimator $%
\widehat{\bolds{\beta}}$, and the correct asymptotic covariance can only
be derived from a set of moment conditions that includes ones for $m_{%
\bolds{\beta}}(X,Z)$ as well as for $\mu$. Conservative inferences again
result if $\widehat{\bolds{\beta}}$ is treated as known when
deriving the
asymptotic covariance matrix.

More generally, for models involving multiple or continuous instrumental
variables, the estimators above are based on unconditional moment conditions
of the form
%
%e10 #&#
\begin{equation}
E \bigl[ \bigl\{ d(Z)-\mu_{d} \bigr\} E(Y_{0}|Z) \bigr]
=0, \label
{GenSMMMC}
\end{equation}
where $E(Y_{0}|Z)$ is determined by the SMM, $d(Z)$ is a user-specified
function, and $\mu_{d}=E\{d(Z)\}$. The choice of $d(Z)$ does not affect
consistency but does affect efficiency. \citet{Rob94} derives the
choice of
$d(Z)=d_{\mathrm{opt}}(Z)$ for the additive and multiplicative SMMs so that the
first-order asymptotitc variance is minimised and the estimator is
semiparametrically efficient; \citet{VanGoe03} derive
the equivalent choice for the double-logistic SMM. For further details
see, for example, \citet{Tsi06} and \citet{BowVan11}.

%s2.4 #&#
\subsection{Covariates}\label{sec2.4}

In this paper we focus mainly on SMMs that do not condition on baseline
covariates, but for completeness we discuss here the estimation of SMMs
which do include covariates; the treatment of covariates is discussed
further in Section~\ref{sec9}. A generalised SMM with baseline covariates
$\mathbf{C}$
has the form
\begin{eqnarray*}
&&h\bigl\{E(Y|X,Z,\mathbf{C})\bigr\}-h\bigl\{E(Y_{0}|X,Z,\mathbf{C})
\bigr\}\\
&&\quad =\eta_{\bolds{\psi}%
}(X,Z,\mathbf{C}),
\end{eqnarray*}
where $\bolds{\psi}$ is the SMM parameter vector and $\eta_{\bolds{\psi
}}(X,Z,\allowbreak \mathbf{C})$ must satisfy $\eta_{\bolds{\psi}}(0,Z,\mathbf{C})=0$.
If $h^{-1}$ is nonseparable, then the association model is specified as
$%
h \{ E(Y|X,Z,\allowbreak\mathbf{C}) \} =m_{\bolds{\beta
}}(X,Z, \mathbf{C)}$.
In terms of identifying assumptions, CMI is now conditional on baseline
$%
\mathbf{C}$ such that
\[
E(Y_{0}|Z,\mathbf{C})=E(Y_{0}|\mathbf{C}),
\]
where NEM corresponds to $\eta_{\bolds{\psi}}(X,Z,\mathbf{C})=\eta
_{%
\bolds{\psi}}(X,\mathbf{C})$ and alternative dimension-reducing parametric
constraints are discussed by\vadjust{\goodbreak} \citet{VanGoe05} and \citet{Tan10}. Finally, the unconditional moment condition (\ref{GenSMMMC}) on
which the estimating equations are based becomes
\[
E \bigl[ \bigl\{d(Z,\mathbf{C})-\mu_{d}(\mathbf{C})\bigr
\}E(Y_{0}|Z,\mathbf {C)} \bigr] =0,
\]
where $E(Y_{0}|Z,\mathbf{C)}$ is determined, as before, by the SMM, $%
E(Y_{0}|Z,\mathbf{C})=d(Z,\mathbf{C})$ is a user-specified function,
and $%
\mu_{d}(\mathbf{C})=E\{d(Z,\mathbf{C})|\mathbf{C}\}$. Consistency thus
depends on correctly specifying the conditional distribution of $Z$
given $%
\mathbf{C}$ so that $\mu_{d}(\mathbf{C})$ is correct for given~$d$. \citet{Rob94} and \citet{VanGoe03} derive the optimal choices
of $d$ for additive, multiplicative and double-logistic SMMs when $\Pr
(Z=z|%
\mathbf{C})$ is presumed to be known; see also \citet{BowVan11}.

An important special case for Mendelian randomisation studies is where there
are discrete baseline covariates to handle population stratification;
see, for example, \citet{Lawetal08}. The generalized SMM with
saturated covariate
effects can be written
\begin{eqnarray*}
&&h\bigl\{E(Y|X,Z,\mathbf{C}=\mathbf{c})\bigr\}-h\bigl\{E(Y_{0}|X,Z,
\mathbf {C}=\mathbf{c}%
)\bigr\}\\
&&\quad =X\psi_{\mathbf{c}},
\end{eqnarray*}
where NEM is taken to hold, and $\psi_{\mathbf{c}}$ is a unique parameter
for the population in the stratum defined by $\mathbf{C}=\mathbf{c}$.
Saturated models of this form are equivalent to specifying separate
no-covariate SMMs within each stratum. Therefore, it can be shown that all
of the results in this paper regarding no-covariate SMMs also apply to
saturated-covariate SMMs; see also \citet{AngImb95}, Theorem 3.

\citet{Tan10} develops an alternative family of doubly robust estimating
equations specifically for generalised SMMs with nonseparable inverse link
functions that include continuous covariates. Furthermore, he allows
for the
inclusion of an extended set of covariates $\mathbf{V}$ that includes $
\mathbf{C}$ so that additional covariates predictive of $Z$, $X$ and
$Y$ can
be incorporated. The analyst first chooses a working distribution
$p^{\ast
}(z|\mathbf{c})$ for $\Pr(Z=z|\mathbf{C}=\mathbf{c})$ that is
arbitrary and
so does not have to be correct. The analyst must then specify two sets of
parametric models involving the full covariates $\mathbf{V}$: (a) $\Pr
(Z=z|%
\mathbf{V}=\mathbf{v})=k_{\lambda}(z|\mathbf{v})$; and (b) $\Pr
(X=x|Z=z,%
\mathbf{V}=\mathbf{v})=g_{\alpha}(x|z,\mathbf{v})$ and
$E(Y|X,Z,\mathbf{V)=}%
m_{\bolds{\upsilon}}^{\ast}(X,Z,\mathbf{V})$. Using the law of iterated
expectations, it can be shown that the following estimating equation is
consistent for $\widehat{\bolds{\psi}}$ if either model (a) or
model (b)
are\vadjust{\goodbreak} misspecified (but not both):
\begin{eqnarray*}
&&n^{-1}\sum_{i} \biggl[ \frac{p^{\ast}(Z_{i}|\mathbf{C}_{i})}{k_{%
\widehat{\lambda}}(Z_{i}|\mathbf{V}_{i})}\phi^{i}
\Delta_{\widehat
{\bolds{\psi}},\widehat{\bolds{\beta}}}^{i}\\
&&\hspace{39pt}{}- \biggl\{ \frac{p^{\ast
}(Z_{i}|\mathbf{C}_{i})}{k_{\widehat{\lambda}}(Z_{i}|\mathbf{V}_{i})}\phi ^{i}
\widehat{w}^{i}-E_{Z}^{\ast} \bigl(
\phi^{i}\widehat{w}^{i} \bigr) \biggr\} \biggr] =0,
\end{eqnarray*}
where\vspace*{2pt} $\Delta_{\bolds{\psi},\bolds{\beta}}^{i}=Y_{i}-h^{-1}\{
m_{\bolds{\beta}}(X_{i},Z_{i},\mathbf{C}_{i})\}+\break h^{-1}  \{ m_{\bolds{\beta}%
}(X_{i}, Z_{i},\mathbf{C}_{i})-\eta_{\bolds{\psi
}}(X_{i},Z_{i},\mathbf{C}%
_{i}) \} $,
\begin{eqnarray*}
\widehat{w}^{i}&=&\sum_{x^{\prime}}g_{\widehat{\alpha}}
\bigl(x^{\prime
}|Z_{i},\mathbf{V}_{i}\bigr) \bigl[
h^{-1} \bigl\{ m_{\widehat{\bolds{\upsilon}}%
}^{\ast}\bigl(x^{\prime},Z_{i},
\mathbf{V}_{i}\bigr) \bigr\}\\
&&\hspace{110pt}{} +\Delta _{\widehat{%
\bolds{\psi}},\widehat{\bolds{\beta}}}^{i}-Y_{i}
\bigr] ,
\end{eqnarray*}
is an estimator of $E ( \Delta_{\psi,\beta}^{i}|Z,\mathbf
{V} ) $, and $E_{Z|\mathbf{C}=\mathbf{c}}^{\ast}(\cdot)=\sum_{z^{\prime
}}p^{\ast}(z^{\prime}|\mathbf{c})(\cdot)$ if $Z$ is discrete. Three important
features to note are that $\Delta_{\widehat{\bolds{\psi}},\widehat
{%
\bolds{\beta}}}^{i}-Y_{i}$ does not depend on $Y_{i}$, $\Delta
_{\bolds{\psi},\bolds{\beta}}^{i}$ is the key to identification because
$E \{
E ( \Delta_{\psi,\beta}^{i}|Z_{i},\mathbf{V}_{i} )
|\mathbf{C}%
_{i} \} =E(Y_{i0}|Z_{i},\mathbf{C}_{i})$, and, while $p^{\ast}$ does
not need to be correctly specified, one must construct $\phi
^{i}=d(Z_{i},
\mathbf{C}_{i})-\mu^{\ast}(\mathbf{C}_{i})$ for user-specified $d$
where $%
\mu^{\ast}(\mathbf{C})=E_{Z|\mathbf{C}}^{\ast} \{ d(Z,\mathbf
{C}%
) \} $. \citet{Tan10} also considers other doubly robust estimating
schemes and argues that the estimator based on the estimating equations
above is locally efficient given the analyst's choices of $p^{\ast}$
and $d$.

%s3 #&#
\section{The Generalised Method of Moments}\label{sec3}

In this section we propose an alternative approach to constructing
estimating equations based on the generalized method of moments (GMM).
\citet{Han82} proposed GMM for moment-condition models of the form
$E (
\mathbf{g} ( \bolds{\delta} )  ) =\mathbf{0}$,
where $%
\mathbf{g} ( \bolds{\delta} ) $ is a random vector and a function
of parameter $\bolds{\delta}$, and $\mathbf{0}$ is an appropriately
dimensioned column vector of zeros. A general expression for the GMM
estimator is given by
%
%e11 #&#
\begin{eqnarray}\label{GMM}
\widehat{\bolds{\delta}}&=&
\mathop{\arg\min}_{\bolds{\delta}}
\Biggl\{
n^{-1}\sum_{i=1}^{n}\mathbf{g}_{i}^{\prime}
( \bolds{\delta} %%
 ) \Biggr\}
 \nonumber\\[-8pt]\\[-8pt]\nonumber
&&{}\cdot   W_{n}^{-1} \Biggl\{
n^{-1}\sum_{i=1}^{n}\mathbf{g}%
_{i}
( \bolds{\delta} ) \Biggr\} ,
\end{eqnarray}
where $\mathbf{g}_{i}(\bolds{\delta})$ is the random vector for
subject $i$, $\mathbf{g}_{i}^{\prime}(\bolds{\delta})$ is its transpose, and $W_{n}$
is a user-chosen weight-matrix that determines the efficiency of the
estimator. \citet{Tan10} has applied the theory of GMM to the doubly robust
estimating equations discussed in the previous section, but the focus here
is on its use in econometrics for instrumental variables models of the form
%
%e12 #&#
\begin{equation}
\mathbf{g} ( \bolds{\delta} ) =v(\bolds{\delta})\mathbf{S}, \label{GenResid}
\end{equation}
where $v(\bolds{\delta)}$ is known as the generalized residual and $%
\mathbf{S}$ is a random vector of instrumental variables. The generalized
residual is so called because it satisfies $E ( v(\bolds{\delta}
)|\mathbf{S}%
 ) =0$. We show how any nonlinear SMM can be expressed as an
instrumental variables model by exploiting that $E \{
Y_{0}-E(Y_{0})|Z \} =0$ under CMI (\ref{cmi}) and by developing
estimating equations which are sample analogues of
\[
E \bigl[ d(\mathbf{S})E \bigl\{ Y_{0}-E(Y_{0})|\mathbf{S}
\bigr\} \bigr] =0,
\]
where $d(\mathbf{S})$ is a user-specified function that affects efficiency
but not consistency. The choice of $d(\mathbf{S})$ that minimises the
variance of the GMM estimator, the so-called efficient instrument, depends
on $W_{n}$ and will be discussed further on.

In our simple scenario involving only binary variables, the SMM is just
identified in the sense that it has one parameter and one moment condition
under CMI (for now taking $\bolds{\beta}$ to be known for the
double-logistic SMM). For example, the additive SMM under NEM leads to the
well-known estimator
%
%e13 #&#
\begin{equation}
\widehat{\psi}_{0}=\frac{E(Y|Z=1)-E(Y|Z=0)}{E(X|Z=1)-E(X|Z=0)}, \label{ClassicalIV}
\end{equation}
in this case, namely, the classical instrumental variable estimator;
see, for example, \citet{HerRob06}. Theory based on the GMM
estimator (\ref{GMM}) is not needed here because $\widehat{\psi}_{0}$ is simply the
solution to (\ref{MCAdd}) under NEM, and the choice of $d(\mathbf
{S})$ is
irrelevant because $Z$ is binary. However, we can use this simple
example to
show how the additive SMM can be specified as an instrumental variables
model.

First, the CMI moment condition can be written as $E(Y_{0}|Z=z)-\alpha
_{0}=0$ for $z=0,1$, where $E(Y_{0})$ is simply treated as an extra
parameter $\alpha_{0}$ and results in the additional moment condition $
E(Y_{0})-\alpha_{0}=0$. It follows that one of $E(Y_{0}|Z=z)-\alpha_{0}=0$
is redundant because $Z$ is discrete and $E\{E(Y_{0}|Z)\}=\alpha_{0}$ by
definition. However, using the additional $E(Y_{0})-\alpha_{0}=0$ moment
condition allows the system of moment conditions to be expressed in
terms of
a generalised residual and a vector of instrumental variables as in
(\ref%
{GenResid}). For example, under the additive SMM, it follows that
%
%e14 #&#
\begin{eqnarray} \label{MCAddNEM}
&&\lleft[ %
\begin{array} {c} E(Y-\psi_{0}X)-
\alpha_{0}
\\
E(Y-\psi_{0}X|Z=1)-\alpha_{0}%
\end{array}
 \rright] =\pmatrix{0\cr 0}\nonumber\\[-8pt]\\[-8pt]
 &&\quad \Rightarrow\quad  E\lleft[ %
\begin{array} {c} Y-\psi_{0}X-\alpha_{0}
\\
(Y-\psi_{0}X-\alpha_{0})Z%
\end{array}
 \rright] =\pmatrix{0\cr 0},\nonumber
\end{eqnarray}
that is, $E\{\mathbf{g}(\psi_{0},\alpha_{0})\}=\mathbf{0}$, where
$\mathbf{%
g}(\psi_{0},\alpha_{0})=(Y-\psi_{0}X-\alpha_{0})\mathbf{S}$ and
$\mathbf{%
S}=(1,Z)^{\prime}$. Similarly, for the multiplicative SMM, it follows
that
%
%e15 #&#
\begin{equation}
E\lleft[ %
\begin{array} {c} Y\exp ( -\psi_{0}X ) -
\alpha_{0}
\\
\bigl\{Y\exp ( -\psi_{0}X ) -\alpha_{0}\bigr\}Z%
\end{array} %
 \rright] =\pmatrix{0\cr 0}, \label{MCMultNEM}
\end{equation}
and for the double-logistic SMM with a saturated association model,
%
%e16 #&#
%%\begin{aligned}
%&&E\lleft[ %
%{\begin{array}{c} \expit (\beta_{0}+
%{}+\beta_{3}XZ-\psi
%_{0}X)-\alpha_{0}
%{}+\beta_{3}XZ-\psi _{0}X)-
% \rright]\nonumber\\[-8pt]\\[-8pt]
% &&\quad  =
%%\end{aligned}
%
\begin{eqnarray}\label{MCDLogisticNEM}
&&E\lleft[\hspace*{-38pt} %
\matrix{ \expit (\beta_{0}+
\beta_{1}X+\beta_{2}Z
\cr
\hspace*{50pt}{}+\beta_{3}XZ-\psi
_{0}X)-\alpha_{0}
\cr
\bigl\{\expit (\beta_{0}+\beta_{1}X+
\beta_{2}Z
\cr
\hspace*{66pt}{}+\beta_{3}XZ-\psi _{0}X)-
\alpha_{0}\bigr\}Z%
} %
 \rright]\nonumber\\[-8pt]\\[-8pt]
 &&\quad  =
\pmatrix{0\cr 0}.\nonumber
\end{eqnarray}

The estimators for these three models are trivial special cases of GMM
because each is just identified, but it is clear that moment conditions
(\ref%
{MCAddNEM})-(\ref{MCMultNEM}) are of the form $E ( v ( \bolds{\delta
} ) \mathbf{S} ) =\mathbf{0}$, where $\mathbf{0}$ is an
appropriately dimensioned vector of zeros. It is also clear that moment
condition (\ref{MCDLogisticNEM}) for the double-logistic SMM has the more
complicated form $E\{\mathbf{g} ( \bolds{\delta};\bolds{\beta}
) \}=%
\mathbf{0}$, because the vector of association model parameters
$\bolds{\beta}$ is usually unknown. We now discuss what happens when $\mathbf{S}$
is expanded to include multiple instrumental variables.

%s4 #&#
\section{Multiple Instruments}\label{sec4}

Mendelian randomisation studies justify the use of genetic markers as
instrumental variables by arguing that (a) the random allocation of genes
from parents to offspring mimics a randomised experiment, and (b)~there
is an
established relationship between the marker and some modifiable risk factor
of interest; see, for example, \citet{Kat86}, \citet{DavEbr03} and
\citet{Lawetal08}.

The genetic variant typically has three forms: homozygous for the common
allele; heterozygous; and homozygous for the rare allele. If we code
these $%
0 $, $1$ and $2$, respectively, then the resulting instrument $Z$ is
multivalued. In fact, this is a simple multiple instruments example because
the three-level variable can be coded using two orthogonal binary variables,
for example, $Z_{1}=I(Z=1)$ and $Z_{2}=I(Z=2)$, where $I$ is the indicator
function.

%s4.1 #&#
\subsection{Additive SMM}\label{sec4.1}

The additive SMM for multiple instruments in this case can be written as
\begin{eqnarray*}
&&E(Y|X,Z_{1},Z_{2})-E(Y_{0}|X,Z_{1},Z_{2})\\
&&\quad =
( \psi_{0}+\psi _{1}Z_{1}+\psi_{2}Z_{2}
) X,
\end{eqnarray*}
where NEM corresponds to constraining $\psi_{1}=\psi_{2}=0$ and CMI
yields the moment conditions
\[
\lleft\{ %
\begin{array} {c} E(Y-\psi_{0}X-
\alpha_{0})
\\
E(Y-\psi_{0}X-\alpha_{0}|Z_{1}=1)
\\
E(Y-\psi_{0}X-\alpha_{0}|Z_{2}=1)%
\end{array} %
 \rright\} =\lleft( %
\begin{array} {c} 0
\\
0
\\
0%
\end{array} %
 \rright) ,
\]
where $\alpha_{0}=E(Y_{0})$ as before. The unconditional moment
condition is
\[
E\bigl\{(Y-\psi_{0}X-\alpha_{0})\mathbf{S}\bigr\}=
\mathbf{0,}
\]
where $\mathbf{S}= ( 1,Z_{1},Z_{2} ) ^{\prime}$ is a
random vector
representing the multiple instruments; note that $\mathbf{S}$ is orthogonal
because its elements are mutually exclusive such that $\mathbf
{SS}^{\prime}=%
\diag(\mathbf{S})$. In fact, this model is linear and so the
parameters can be consistently estimated using standard Two-Stage Least
Squares (2SLS). The 2SLS estimator can be obtained as the ordinary least
squares (OLS) estimator from regressing $Y$ on $\widehat{X}$, where $%
\widehat{X}$ is the prediction from the first-stage regression of $X$
on~$\mathbf{S}$. The 2SLS estimator is a special case of a ``one-step'' GMM
estimator with $W_{n}=n^{-1}\sum_{i}\mathbf{S}_{i}\mathbf
{S}_{i}^{\prime}$
(see next section), and is commonly used for linear instrumental variables
analysis with multiple instruments; see \citet{Paletal12} for its use
with Mendelian randomisation studies.

%s4.2 #&#
\subsection{Multiplicative SMM}\label{sec4.2}

The saturated multiplicative SMM for the two instruments is
\begin{eqnarray*}
&&E(Y|X,Z_{1},Z_{2})/E(Y_{0}|X,Z_{1},Z_{2})\\
&&\quad =
\exp\bigl\{ ( \psi_{0}+\psi _{1}Z_{1}+
\psi_{2}Z_{2} ) X\bigr\},
\end{eqnarray*}
where NEM here corresponds to $\psi_{1}=\psi_{2}=0$. Using the same
vector of instrumental variables $\mathbf{S}$, the multiplicative SMM moment
conditions can be written as
%
%e17 #&#
\begin{equation}
E \biggl[ \biggl\{ \frac{Y}{\exp ( X\psi_{0} ) }-\alpha _{0} \biggr\} \mathbf{S}
\biggr] =\mathbf{0}. \label{mmom0}
\end{equation}
Letting $\alpha_{0}^{\ast}=\log(\alpha_{0})$, it is easy to show
that (%
\ref{mmom0}) also implies
%
%e18 #&#
\begin{equation}
E \biggl\{ \frac{Y-\exp ( \alpha_{0}^{\ast}+X\psi_{0} )
}{\exp
 ( X\psi_{0} ) }\mathbf{S} \biggr\} =\mathbf{0} \label{mmom1}
\end{equation}
and
%
%e19 #&#
\begin{equation}
E \biggl\{ \frac{Y-\exp ( \alpha_{0}^{\ast}+X\psi_{0} )
}{\exp
 ( \alpha_{0}^{\ast}+X\psi_{0} ) }\mathbf{S} \biggr\} =\mathbf{0}, \label{mmomc}
\end{equation}
where (\ref{mmomc}) is obtained simply by dividing (\ref{mmom1}) by
$\exp
(\alpha_{0}^{\ast})\neq0$. Moment condition (\ref{mmomc}) is the
same as
that for exponential-mean models proposed by \citet{Mul97}.

For example, consider a GMM estimator based on moment condition (\ref
{mmom0}%
). The GMM estimator for $\bolds{\delta}= ( \alpha_{0},\psi
_{0} ) ^{\prime}$ is the solution to (\ref{GMM}) with $\mathbf
{g}%
 ( \bolds{\delta} ) =\{Y\exp ( -X\psi_{0} )
-\alpha
_{0}\}\mathbf{S}$. The one-step GMM estimator\vspace*{2pt} $\widehat{\bolds{\delta}}%
_{1} $ is obtained by choosing the weight matrix in (\ref{GMM}) to be $
W_{n}=n^{-1}\sum_{i}\mathbf{S}_{i}\mathbf{S}_{i}^{\prime}$. The two-step
GMM estimator $\widehat{\bolds{\delta}}_{2}$ is obtained by estimating
the weight matrix
\[
W_{n} ( \widehat{\bolds{\delta}}_{1} ) =n^{-1}
\sum_{i=1}^{n}\mathbf{g}_{i} ( \widehat{\bolds{\delta }%
}_{1} ) \mathbf{g}_{i}^{\prime} (
\widehat{\bolds{\delta}}%
_{1} ) ,
\]
using the one-step GMM estimator $\widehat{\bolds{\delta}}_{1}$. Under
standard regularity conditions, the limiting distributions of the one-step
and two-step GMM estimators are
\begin{eqnarray*}
n^{1/2} ( \widehat{\bolds{\delta}}_{1}-\bolds{\delta
}_{0} ) &\stackrel{d} {\longrightarrow}&N \bigl\{ \mathbf{0}, \bigl(
C_{0}^{\prime
}W^{-1}C_{0} \bigr)
^{-1}C_{0}W^{-1}\\
&&\hspace*{26pt}{}\cdot\Omega_{0}W^{-1}C_{0}
\bigl( C_{0}^{\prime}W^{-1}C_{0} \bigr)
^{-1} \bigr\},
\\
n^{1/2} ( \widehat{\bolds{\delta}}_{2}-\bolds{\delta
}_{0} )&\stackrel{d} {\longrightarrow}&N \bigl\{ \mathbf{0}, \bigl(
C_{0}^{\prime
}\Omega_{0}^{-1}C_{0}
\bigr) ^{-1} \bigr\} ,
\end{eqnarray*}
respectively, where $\bolds{\delta}_{0}$ is the true parameter
value, $%
\stackrel{d}{\longrightarrow}$ indicates convergence in distribution, $N$
indicates a normally distributed random vector,
\[
C_{0}=E \biggl\{ \frac{\partial\mathbf{g} ( \bolds{\delta
}_{0} ) }{%
\partial\bolds{\delta}^{\prime}} \biggr\} ,\quad \Omega_{0}=E
\bigl\{ \mathbf{g}%
 ( \bolds{\delta}_{0} )
\mathbf{g}^{\prime} ( \bolds{\delta }_{0} ) \bigr\} ,
\]
and $W=E(\mathbf{S}_{i}\mathbf{S}_{i}^{\prime})$ is the probability limit
of the one-step GMM estimator's weight matrix.

\citet{Cha87} shows that the two-step GMM estimator is
semiparametrically efficient when the instruments are mutually exclusive
indicators that follow a multinomial distribution, as is the case in this
example provided that there are no continuous covariates or instruments.
More generally, as will be discussed in Section~\ref{sec5}, one must derive the
efficient instrument $d(\mathbf{S})=d_{\mathrm{opt}}(\mathbf{S})$ for the GMM
estimator to be semiparametrically efficient.

A useful property of two-step GMM for over-identified models is that it
admits the use of the Hansen \textit{J}-test, which can be used to assess
the validity of the moment conditions; see \citet{Han82}. The test statistic
and its limiting distribution (under the null hypothesis that the moment
conditions are valid) are given by
\begin{eqnarray*}
J ( \widehat{\bolds{\delta}}_{2} )& =&n \Biggl\{ n^{-1}
\sum_{i=1}^{n}\mathbf{g}_{i}^{\prime} (
\widehat{%
\bolds{\delta}}_{2} ) \Biggr\}\\
&&{}\cdot W_{n}^{-1}
( \widehat {\bolds{\delta}}_{1} ) \Biggl\{ n^{-1}
\sum_{i=1}^{n}\mathbf{g}%
_{i} ( \widehat{
\bolds{\delta}}_{2} ) \Biggr\}\\
&\stackrel  {d} {%
\longrightarrow}&\chi_{q}^{2},
\end{eqnarray*}
where $\chi_{q}^{2}$ indicates a chi-squared random variable with $q$
degrees of freedom, and $q$ is the number of moment conditions by which the
model is over identified (e.g., $q=1$ in this illustration).

%s4.3 #&#
\subsection{Double-Logistic SMM}\label{sec4.3}

Under NEM, the logistic SMM for the two instruments is
\begin{eqnarray*}
&&\logit \bigl\{E(Y|X,Z_{1},Z_{2})\bigr\}-\logit
\bigl\{ E(Y_{0}|X,Z_{1},Z_{2})\bigr\}\\
&&\quad =%
\psi_{0}X,
\end{eqnarray*}
and its association model is
%
%e20 #&#
\begin{equation}
E(Y|X,Z_{1},Z_{2})=\expit  \bigl\{
m_{\bolds{\beta
}}(X,Z_{1},Z_{2})%
 \bigr\} ,
\label{SatLog}
\end{equation}
where $m_{\bolds{\beta}}(X,Z_{1},Z_{2})=\beta_{0}+\beta_{1}X+\beta
_{2}Z_{1}+\beta_{3}Z_{2}+\beta_{4}XZ_{1}+\beta_{5}XZ_{2}$ is saturated.
We describe two estimation methods: first, where the parameters in the
saturated association model are estimated by maximum likelihood and then
plugged into the estimating equations for the double-logistic SMM; and,
second, where all parameters are estimated jointly in a similar manner to
that proposed by \citet{VanGoe03} and \citet{BowVan11}.

Denoting $\widehat{\bolds{\beta}}$ as the maximum likelihood
estimator of~$\bolds{\beta}$, it follows that
%
%e21 #&#
\begin{equation}
E\bigl\{\mathbf{g}(\bolds{\delta};\widehat{\bolds{\beta}})\bigr\}=E\bigl[\bigl
\{ q(\psi_{0};%
\widehat{\bolds{\beta}})-
\alpha_{0}\bigr\}\mathbf{S\bigr]}=\mathbf{0} , \label{2SGMM}
\end{equation}
where $\bolds{\delta}=(\psi_{0},\alpha_{0})^{\prime}$, $q(\psi
_{0};%
\bolds{\beta})=\expit \{m_{\bolds{\beta}} (
X,Z_{1},\allowbreak Z_{2} ) -X\psi_{0}\}$ and $\mathbf
{S}=(1,Z_{1},Z_{2})^{\prime
} $. Point estimation is carried out exactly as before, but standard error
estimates obtained by fixing $\widehat{\bolds{\beta}}$ and plugging it
into the asymptotic covariance matrices presented above will be biased
because the first stage estimation of $\bolds{\beta}$ is ignored;
see the
discussion in Section~\ref{sec2.3}. However, theory for ``two-stage'' GMM estimators
(2SGMM) has been developed by \citet{GouMonRen96}. The 2SGMM $%
\widehat{\bolds{\delta}}_{1,\bolds{\beta}}$ is the solution to
(\ref{GMM}) and its asymptotic distribution is
\begin{eqnarray*}
&&n^{1/2} ( \widehat{\bolds{\delta}}_{1,\bolds{\beta
}}-\bolds{\delta}%
_{0} )\\
&&\quad  \stackrel {d} {\longrightarrow}N \bigl\{
\mathbf{0}, \bigl( C_{0}^{\prime}WC_{0} \bigr)
^{-1}C_{0}W\Omega_{0}^{\ast}WC_{0}
\bigl( C_{0}^{\prime}WC_{0} \bigr) ^{-1}
\bigr\} ,
\end{eqnarray*}
where $C_{0}$ and $W$ are both defined as above, and $\Omega_{0}^{\ast}$
is the asymptotic variance of the limiting normal distribution of
\[
n^{-1/2}\sum_{i=1}^{n}\mathbf{g}_{i} (
\bolds{\delta}_{0};%
\bolds{\beta}_{0} ) +E
\biggl\{ \frac{\partial\mathbf
{g} ( \bolds{\delta}_{0};\bolds{\beta}_{0} ) }{\partial\bolds{\beta
}^{\prime}}%
 \biggr\} n^{1/2} ( \widehat{\bolds{\beta}}-\bolds{\beta }_{0} ) ,
\]
which has the consistent estimator
\begin{eqnarray*}
n\widehat{\Omega}^{\ast}&=&\sum_{i=1}^{n}\widehat{
\mathbf{g}}_{i}%
\widehat{\mathbf{g}}_{i}^{\prime}+
\widehat{G}_{\bolds{\beta
}}^{\prime}%
\widehat{V(\widehat{\bolds{\beta}})}\widehat{G}_{\bolds{\beta
}}\\
&&{}+\widehat{%
G}_{\bolds{\beta}}^{\prime}
\widehat{V(\widehat{\bolds{\beta }})} \Biggl( \sum_{i=1}^{n}Q_{i}
\mathbf{R}_{i}\widehat{\mathbf{g}}_{i}^{\prime
} \Biggr)\\
&&{} +
\Biggl( \sum_{i=1}^{n}Q_{i}\widehat{
\mathbf{g}}_{i}%
\mathbf{R}_{i}^{\prime} \Biggr)
\widehat{V(\widehat{\bolds{\beta }})}%
\widehat{G}_{\bolds{\beta}},
\end{eqnarray*}
with $\widehat{\mathbf{g}}_{i}=\mathbf{g}_{i}(\widehat{\bolds{\delta}}_{1,%
\bolds{\beta}};\widehat{\bolds{\beta}})$, $\widehat{G}_{\bolds{\beta}%
}=\sum_{i}\partial\mathbf{g}_{i}^{\prime}(\widehat{\bolds{\delta
}}_{1,%
\bolds{\beta}};\widehat{\bolds{\beta}})/\partial\bolds{\beta
}$,  $%
\widehat{V(\widehat{\bolds{\beta}})}= ( \sum_{i}\widehat
{p}_{i} (
1-\widehat{p}_{i} ) \mathbf{R}_{i}\mathbf{R}_{i}^{\prime
} ) ^{-1}$%
, $\mathbf{R}_{i}= (
1,X_{i},Z_{1i},\allowbreak Z_{2i},X_{i}Z_{1i},X_{i}Z_{2i} ) ^{\prime}$,
$\widehat{p%
}_{i}=\expit \{m_{\widehat{\bolds{\beta
}}}(X_{i},Z_{i1},Z_{i2})\}$
and $Q_{i}=Y_{i}-\widehat{p}_{i}$. Furthermore, $\widehat{\Omega
}^{\ast}$
is also the weight matrix for the asymptotically efficient two-step 2SGMM
estimator, and so the limiting distribution of the Hansen \textit{J}-test
statistic (with $W_{n}=\widehat{\Omega}^{\ast}$) is also valid.

\citet{VanGoe03} developed estimating equations for the
double-logistic SMM by expanding its system of estimating equations to
include those for the association model. As in \citet{BowVan11}, a joint GMM estimator can be obtained by applying the GMM estimator
to
%
%e22 #&#
\begin{eqnarray}\label{lmomjoint}
 \qquad &&\mathbf{g} ( \bolds{\delta};\bolds{\beta} )\nonumber\\[-8pt]\\[-8pt]
 &&\quad  =\lleft( %
\begin{array} {c}
{}\bigl[ Y-\expit  \bigl\{ m_{\bolds{\beta}} ( X,Z_{1},Z_{2}
) \bigr\} \bigr]\mathbf{R}
\\
{}\bigl[\expit \big\{m_{\bolds{\beta}} ( X,Z_{1},Z_{2} ) -
\psi _{0}X-\alpha_{0}\bigr\}\bigr]\mathbf{S}%
\end{array}
 \rright) ,\nonumber
\end{eqnarray}
where $\mathbf{R}$ is defined above and $\bolds{\delta}= (
\alpha
_{0},\psi_{0} ) ^{\prime}$. Gouri{\'e}\-roux, Monfort and Renault (\citeyear{GouMonRen96}) show that
the asymptotic distributions of the 2SGMM and the joint GMM estimators are
the same. An important advantage of using the joint moments (\ref
{lmomjoint}%
) is that standard GMM software can be used to make asymptotically correct
inferences about the target parameter $\psi_{0}$. Further details on how
the \textit{gmm} command in Stata and the \textit{gmm}() function in
R can
be used to implement these estimators are given in the \hyperref[app]{Appendix}.

%s5 #&#
\section{Combining Multiple Instruments}\label{sec5}

\citet{BowVan11} derive the optimally efficient
combination of
instruments and, for practical purposes, a simplified expression for this
combination. We consider the particular case of SMMs without covariates
where identification is obtained using orthogonal binary instruments. In
such cases, we show that the one-step GMM estimator combines the instruments
as in \citet{BowVan11} under the simplifying assumption
of a
constant variance, and that the two-step GMM estimator combines the
instruments optimally.

First consider the one-step GMM estimator by noting that it is the solution
to the first derivative of (\ref{GMM}) evaluated at zero. For the
multiplicative SMM based on (\ref{mmom0}), this gives
\begin{eqnarray*}
&& \Biggl\{ n^{-1}\sum_{i=1}^{n}
\frac{\partial\mathbf{g}%
_{i}^{\prime} ( \bolds{\delta} ) }{\partial\bolds{\delta}}%
 \Biggr\} W_{n}^{-1} \Biggl\{
n^{-1}\sum_{i=1}^{n}\mathbf{g}%
_{i}
( \bolds{\delta} ) \Biggr\}
\\
&&\quad =\lleft\{ n^{-1}\sum_{i=1}^{n}\lleft(
\begin{array} {c} 1
\\
Y_{i}X_{i}\exp ( -X_{i}\psi_{0}
)%
\end{array} %
 \rright) \mathbf{S}_{i}^{\prime}
\rright\}\\
&&\quad \quad {}\cdot W_{n}^{-1} \Biggl\{ n^{-1}
\sum_{i=1}^{n}\mathbf{g}_{i} ( \bolds{\delta} )
\Biggr\} =\mathbf{0},
\end{eqnarray*}
where $\mathbf{g}_{i} ( \bolds{\delta} ) =\{Y\exp
(-X\psi
_{0})-\alpha_{0}\}\mathbf{S}$. This system can be expressed as
\[
B^{\prime}S \bigl( S^{\prime}S \bigr) ^{-1}S^{\prime}
\mathbf {v}=\mathbf{0},
\]
where $B= \{ \mathbf{b}_{i}^{\prime} \} $ and $S= \{
\mathbf{S}%
_{i}^{\prime} \} $ are the matrices formed by stacking the
vectors $%
\mathbf{b}_{i}^{\prime}=(1,Y_{i}X_{i}\*\exp ( -X_{i}\psi
_{0} ) )$
and $\mathbf{S}_{i}^{\prime}$, respectively, and $\mathbf{v}= \{
v_{i} \} $ is a column vector with elements given by
$v_{i}=Y_{i}\exp
 ( -X_{i}\psi_{0} ) -\alpha_{0}$. It is thus apparent that the
GMM estimator combines the instruments in the projection $S (
S^{\prime
}S ) ^{-1}S^{\prime}B$, that is, the multiple instruments for each
individual are replaced by the linear projection of $\mathbf{b}_{i}$ onto
the space spanned by $S$; alternatively put, the combined instrumental
variable can be thought of as the prediction from a linear regression
of $%
\mathbf{b}_{i}$ on the instruments $\mathbf{S}_{i}$.

For the binary variables case considered here, we have that
%
%e23 #&#
\begin{equation}
Y_{i}X_{i}\exp ( -X_{i}\psi_{0} )
=Y_{i}X_{i}\exp ( -\psi _{0} ), \label{gmmproj}
\end{equation}
so that the one-step GMM can be thought of combing the instruments simply
using the linear projection of $YX$ onto the space spanned by $\mathbf{S}$.
The one-step GMM estimator for the double-logistic SMM estimator also has
the form of a linear projection of $\mathbf{b}_{i}$ onto the space spanned
by $S$, but here $\mathbf{b}_{i}^{\prime}= ( 1,q_{i}(\psi
_{0};\widehat{%
\bolds{\beta}}) \{ 1-q_{i}(\psi_{0};\widehat{\bolds{\beta}}
) \} X_{i} ) $. For both the multiplicative and logistic SMMs,
these are the simplified combinations of multiple instruments of \citet{BowVan11}.

In the simple setup involving only binary variables, the one-step GMM
estimator for the multiplicative SMM can be expressed as a linear 2SLS
estimator. Following \citet{Ang01}, note that $\exp ( -\psi
_{0}X ) = ( 1-X ) +X\exp ( -\psi_{0} ) $ and,
therefore,
\[
Y\exp ( -\psi_{0}X ) -\alpha_{0}=Y ( 1-X ) +YX\exp ( -
\psi_{0} ) -\alpha_{0}.
\]
Hence, the moment conditions can be expressed as the linear [in $\exp
 (
-\psi_{0} ) $] moments
%
%e24 #&#
\begin{equation}
\quad E\bigl[ \bigl\{ Y ( 1-X ) +YX\exp ( -\psi_{0} ) -\alpha
_{0} \bigr\} \mathbf{S}\bigr]=\mathbf{0,} \label{lin1}
\end{equation}
from which we see that the one-step GMM estimator for $\exp (
-\psi
_{0} ) $ using moment condition (\ref{mmom0}) is identical to
the 2SLS
estimator from regressing $Y ( X-1 ) $ on $\widehat{YX}$,
where $%
\widehat{YX}$ are the predictions from the linear regression of $YX$
on $S$.

Multiplying (\ref{lin1}) by the risk ratio $\exp ( \psi
_{0} ) $,
we obtain
%
%e25 #&#
\begin{equation}
E\bigl[ \bigl\{ YX+Y ( 1-X ) \exp ( \psi_{0} ) -\gamma
_{0} \bigr\} \mathbf{S\bigr]}=\mathbf{0}, \label{lin2}
\end{equation}
where $\gamma_{0}=\alpha_{0}\exp ( \psi_{0} ) $. In this case,
the same estimator as the one-step GMM estimator for $\exp ( \psi
_{0} ) $ is obtained from a linear instrumental variable estimator
where $ ( X-1 ) Y$ is instrumented by $\widehat{YX}$. We
will use
this result later in Section~\ref{sec6} when deriving results for local risk ratios.

We now move on to the optimal combination of instruments. As we
discussed at
the end of Section~\ref{sec4.2}, \citet{Cha87} established efficiency results
for GMM estimators. We describe these results in terms of a simple
multiplicative SMM and its three moment conditions
%
%e26 #&#
\begin{equation}
E \bigl\{ Y\exp ( -X\psi_{0} ) -\alpha_{0}|Z=z \bigr\} =0,
\label{restr}
\end{equation}
for $z=0,1,2$. As shown previously, the instruments can be represented by
the vector of orthogonal binary instruments $\mathbf{S}$ and the generalized
residual
\[
\nu ( Y,X;\delta_{0} ) =Y\exp ( -X\psi_{0} ) -\alpha
_{0},
\]
where $\delta_{0}=(\alpha_{0},\psi_{0})^{\prime}$. Using the
notation of
\citet{New93}, the efficient instrument is
%
%e27 #&#
\begin{equation}
d_{\mathrm{opt}} ( \mathbf{S} ) =Qe ( \mathbf{S} ) /\sigma ^{2} (
\mathbf{S} ), \label{minvarz}
\end{equation}
where $Q$ is any nonsingular matrix,
\begin{eqnarray*}
e ( \mathbf{S} ) &=&E \biggl\{ \frac{\partial\nu (
Y,X;\delta
_{0} ) }{\partial\delta}\Big|\mathbf{S} \biggr\}\\
& =&-\lleft( %
\begin{array} {c} 1
\\
E \bigl\{ Y\exp ( -X\psi_{0} ) X|\mathbf{S} \bigr\}%
\end{array} %
 \rright) ,
\\
\sigma^{2} ( \mathbf{S} ) &=&E \bigl\{ \nu^{2} ( Y,X;
\delta _{0} ) |\mathbf{S} \bigr\}\\
& =&E \bigl[ \bigl\{ Y\exp ( -X\psi
_{0} ) -\alpha_{0} \bigr\} ^{2}|\mathbf{S}
\bigr] ,
\end{eqnarray*}
which leads to a GMM estimator with asymptotic covariance
\[
\Lambda= \bigl[ E \bigl\{ e ( \mathbf{S} ) e ( \mathbf{S} ) ^{\prime}/
\sigma^{2} ( \mathbf{S} ) \bigr\} \bigr] ^{-1}.
\]
\citet{Cha87} showed that, when $\mathbf{S}$ comprises multinomially
distributed multiple orthogonal binary instruments such that $\mathbf
{SS}%
^{\prime}=\diag(\mathbf{S})$, the asymptotic\vspace*{1pt} covariance of the
two-step GMM $(C_{0}^{\prime}\Omega_{0}^{-1}C_{0})^{-1}=\Lambda$.
Hence, we can derive the optimum combination of instrumental variables from
the first-order condition for the two-step GMM estimator:
\begin{eqnarray*}
&& \Biggl\{ \sum_{i=1}^{n}\frac{\partial\mathbf{g}_{i}^{\prime
} ( \bolds{\delta} ) }{\partial\bolds{\delta}} \Biggr\}
W_{n}^{-1} ( \widehat{\delta}_{1} ) \Biggl\{
\sum_{i=1}^{n}\mathbf{g}_{i} ( \bolds{\delta} )
\Biggr\}
\\
&&\quad =\lleft\{ \sum_{i=1}^{n}\lleft( %
\begin{array} {c} 1
\\
Y_{i}\exp ( -X_{i}\psi_{0} ) X_{i}%
\end{array} %
 \rright) \mathbf{S}_{i}^{\prime}
\rright\}\\
&&\qquad {}\cdot W_{n}^{-1} ( \widehat {\delta}%
_{1}
) \Biggl\{ \sum_{i=1}^{n}\mathbf{g}_{i} (
\bolds{\delta} ) \Biggr\} =\mathbf{0},
\end{eqnarray*}
where $\mathbf{g}_{i} ( \bolds{\delta} ) =\{Y_{i}\exp
(-X_{i}\psi
_{0})-\alpha_{0}\}\mathbf{S}_{i}$, $\mathbf
{S}_{i}=(Z_{io},\allowbreak Z_{i1},Z_{i2})^{%
\prime}$ with $Z_{ij}=I(Z_{i}=j)$, and
\begin{eqnarray*}
&&\hspace*{-6pt}W_{n} ( \widehat{\delta}_{1} )\\
 &&\hspace*{-6pt}\quad =\sum
_{i=1}^{n} \bigl( Y_{i}\exp
(-X_{i}\widehat{\psi}_{1})-\widehat{\alpha}_{1}
\bigr) ^{2}\mathbf {S}_{i}%
\mathbf{S}_{i}^{\prime}
\\
&&\hspace*{-6pt}\quad =\lleft[ %
\begin{array} {cc} \displaystyle\sum
_{i}Z_{i0}\nu^{2} ( Y_{i},X_{i}; \widehat{\delta}_{1} )  &\quad 0
\\
0 & \quad \displaystyle\sum_{i}Z_{i1}\nu^{2} (  Y_{i},X_{i};\widehat{\delta }_{1} )
\\
0 &\quad  0
\end{array} %
 \rright.
 \\
&&\hspace*{-6pt}\quad \hspace*{123pt} \lleft. %
\begin{array} {c}  0
\\
 0
\\
\displaystyle\sum_{i}Z_{i2}v^{2} (
Y_{i},X_{i};\widehat{\delta }_{1} )%
\end{array} %
 \rright] .
\end{eqnarray*}
As before, let the matrices $S$ and $B$ be defined as $B= \{
\mathbf{b}%
_{i}^{\prime} \} $ and $S= \{ \mathbf{S}_{i}^{\prime
} \} $,
obtained by stacking the vectors $\mathbf{b}_{i}^{\prime
}=(1,Y_{i}\exp
 ( -X_{i}\psi_{0} ) X_{i})$ and $\mathbf{S}_{i}^{\prime}$,
respectively, then the way the two-step GMM estimator combines the multiple
instruments is given by
\begin{eqnarray*}
&&SW_{n}^{-1} ( \widehat{\delta}_{1} )
S^{\prime}B
\\
&&\quad =S\diag  \Biggl( \frac{1}{n_{z}}\sum_{i,Z_{i}=z}v^{2}
( Y_{i},X_{i};\widehat{\delta}_{1} ) \Biggr)
^{-1} \bigl( S^{\prime
}S \bigr) ^{-1}S^{\prime}B,
\end{eqnarray*}
which is a consistent estimate for the optimal instruments. \citet{Cha87} further showed that $\Lambda$ is also the lower bound for the
asymptotic variance of any consistent asymptotically normally distributed
estimator of a semiparametric model where the only substantive restriction
imposed on the distribution of the data is (\ref{restr}).

%s6 #&#
\section{Monte Carlo Studies}\label{sec6}

%s6.1 #&#
\subsection{Multiplicative SMM}\label{sec6.1}

We now present two Monte Carlo simulation studies to demonstrate the
properties of GMM estimators with multiple orthogonal binary
instruments in
models without covariates. First, we consider the multiplicative SMM by
generating data from population model~$M_{1}$, which satisfies the
multiplicative SMM under both the NEM and CMI restrictions. Population
model $M_{1}$ is defined so that
\begin{eqnarray*}
E ( Y|X,Z_{1},Z_{2} ) &=&\exp\bigl\{\beta_{0}+ (
\beta _{1}+\psi _{0} ) X+\beta_{2}Z_{1}\\
&&\hspace*{17pt}{}+
\beta_{3}Z_{2}+\beta_{4}XZ_{1}+\beta
_{5}XZ_{2}\bigr\},
\end{eqnarray*}
where $\psi_{0}=0.6$ is the treatment effect. To define the
distribution of
the observed data, we further define $Z$ to follow the marginal distribution
given by $P ( Z=1 ) =0.3$ and $P ( Z=2 ) =0.2$,
and $%
P ( X=1|Z=z ) =p_{10}+0.15\times z$ for $z=0,1,2$. To define the
joint distribution of the observed and potential outcomes, we set the
expected treatment-free outcome in the population to be $\alpha
_{0}=E (
Y_{0} ) =0.19$, which leads to $\alpha_{0}^{\ast}=\log E (
Y_{0} ) =-1.6607$ in moment conditions (\ref{mmom1}) and (\ref{mmomc}),
and $E ( Y ) =0.25$, $\beta_{1}=0.15$, $\beta_{4}=0.6$ and
$\beta
_{5}=-0.6$. The other parameter values are then numerically found in order
for CMI and NEM to hold: $\beta_{0}=-1.6976$, $\beta_{2}=-0.3186$,
$\beta
_{3}=0.2511$ and \mbox{$p_{10}=0.2321$}.

Table~\ref{tab1} presents some estimation results for $10\mbox{,}000$ samples of size $%
10\mbox{,}000 $ drawn from population mo\-del~$M_{1}$. Three different versions of
the GMM estimator are applied: the first column of Table~\ref{tab1} contains the
results of the just-identified model using the multivalued instrument
$Z\in
\{0,1,2\}$ as a single instrument so that $\mathbf{S}=(1,Z)^{\prime
}$; in
the second and third columns, we present the one- and two-step GMM estimates
for moment conditions (\ref{mmom1}) and (\ref{mmomc}), respectively, using
multiple instruments so that $\mathbf{S}=(1,Z_{1},Z_{2})^{\prime}$.

%t1 #&#
\begin{table}
\tabcolsep=3pt
\caption{Monte Carlo estimation results for multiplicative SMM}\label{tab1}
\begin{tabular*}{\columnwidth}{@{\extracolsep{\fill}}ld{2.5}d{2.5}d{2.5}@{}}
\hline
 & \multicolumn{1}{c}{\textbf{Single instrument}} & \multicolumn{2}{c@{}}{\textbf{Multiple instruments}}  \\
\hline
\textbf{Instruments} $\mathbf{S}$ &  \multicolumn{1}{c}{$\bolds{1,Z}$} & \multicolumn{2}{c@{}}{$\bolds{1,Z_{1},Z_{2}}$}  \\
\hline
\textbf{Moment conditions} &  \multicolumn{1}{c}{\textbf{(\ref{mmom1}) or (\ref{mmomc})}} & \multicolumn{1}{c}{\textbf{(\ref{mmom1})}}& \multicolumn{1}{c}{\textbf{(\ref{mmomc})}    }\\
 \hline
One-step GMM &   & &   \\
$\alpha_{0}^{\ast}$  & -1.6614 & -1.6628 & -1.6599  \\
 & (0.0839) & (0.0561) & (0.0561)  \\
 & [0.0843] & [0.0566] & [0.0565]  \\
$\psi_{0}$  & 0.6151 & 0.6102 & 0.6033  \\
 & (0.2175) & (0.1358) &  (0.1353)  \\
 & [0.2168] & [0.1361] &   [0.1356]  \\
Two-step GMM  &  & &  \\
$\alpha_{0}^{\ast}$ &  & -1.6629 & -1.6598  \\
 & & (0.0561) & (0.0561)  \\
& & [0.0565] & [0.0565]  \\
$\psi_{0}$ &  & 0.6095 & 0.6024  \\
 & & (0.1355) & (0.1350)  \\
 & & [0.1359] & [0.1353]  \\
Hansen \textit{J}  &  & 0.9806 & 0.9793  \\
Rej. freq. 5\% &  & 0.0478 & 0.0475  \\
\hline
\end{tabular*}
\tabnotetext[]{tz}{Notes: Sample size 10,000; means based on 10,000
Monte Carlo replications; std. error in brackets; means of estimated
standard errors in square brackets; data drawn from population model $M_{1}$
 as described in Section~\ref{sec6.1}; $\alpha_{0}^{\ast}=-1.6607$
and $\psi_{0}=0.6$.}
\end{table}

All of the estimators display a small positive bias for $\psi
_{0}=0.6$, and
the mean estimated standard errors are very close to the true standard
errors. Among the two estimators using multiple instruments, this bias is
slightly larger for the estimator based on moment condition (\ref{mmom1}).
There is here a negligible gain in precision from using the two-step GMM
estimator as compared to the one-step estimator. However, there is a
substantial gain in efficiency from using two instrumental variables rather
than one, with the standard error decreasing from 0.22 for the
just-identified model to 0.14 for the two-step GMM estimators. This is
because the GMM projection (\ref{gmmproj}) in this case is not linear
in $Z$, even though the conditional probabilities $P ( X=1|Z ) $ are.
More specifically, the coefficient on $Z_{2}$ in the regression of $YX$
on $%
(1,Z_{1},Z_{2})$ from (\ref{gmmproj}) is actually smaller than that of
$%
Z_{1} $. Under this particular population model (but not generally) the
relationship between the coefficients is roughly linear: the average
coefficient on $Z_{1}$ is equal to $0.1067$ and for $Z_{2}$ it equals $%
0.0557 $. Hence, a single instrument that takes the value $1$ if $Z=2$
and $%
2 $ if $Z=1$ leads to a just-identified estimator which is likely to be
almost as efficient as the over-identified GMM estimators. Further
simulations show that this is indeed the case, with the just-identified
estimator for $\psi_{0}$ just described having an average of $0.6077$
and a
standard error of $0.1375$, which are both virtually identical to those of
the over-identified GMM estimators.

We repeated the analysis above for a similar design to $M_{1}$ but with the
instrument $Z$ taking the six values $0,1,\ldots,5$; full details of this
design are available from the authors. The GMM estimators are again well
behaved. Using moment conditions (\ref{mmomc}), the mean based on 10,000
Monte Carlo estimates using the two-step GMM estimator is 0.5966 with a
standard error 0.0801; the mean estimated standard error equals 0.0806.
The~rejection frequency of the \textit{J}-test is 5.1\% at the 5\% level.

Returning to the design with $Z$ taking the values $0,1,2$, we modify
population model $M_{1}$ so as to study how the multiplicative GMM performs
when $Z$ does not satisfy the key conditions of an instrumental
variable. We
do this by keeping all $M_{1}$ parameters the same but making the
``instrument'' $Z_{1}$ invalid. This is
done by specifying
\begin{eqnarray*}
&&E(Y|X,Z_{1},Z_{2})\\
&&\quad =\exp\bigl\{\beta_{0}+ (
\beta_{1}+\psi_{0} ) X+ ( \beta_{2}+\phi )
Z_{1}\\
&&\hspace*{58pt}{}+\beta_{3}Z_{2}+\beta _{4}XZ_{1}+
\beta_{5}XZ_{2}\bigr\},
\end{eqnarray*}
with $\phi=0.15$. In this case, the CMI assumption is violated as
$E [
Y_{0}|Z=0 ] =E [ Y_{0}|Z=2 ] =0.19$ as before, but now
$E [
Y_{0}|Z=1 ] =0.2207$. The GMM estimators are now severely biased
upwards. The mean based on 10,000 Monte Carlo estimates of the two-step GMM
estimator using moments (\ref{mmomc}) is equal to 1.1191, with a standard
error of 0.1681. The mean (variance) of Hansen's \textit{J}-test is
equal to
3.56 (3.70) with a rejection frequency at the 5\% level of 34\%. If instead
we change the coefficient on $Z_{2}$ to $\beta_{3}+0.15$, we get a much
smaller bias, with the mean (std. error) of the estimator equal to 0.6452
(0.1370), but the rejection frequency of the \textit{J}-test is now much
larger, namely, 93\% at the 5\% level. This difference is due to the fact
that, as highlighted above, in this case $Z_{1}$ is a stronger instrument
than $Z_{2}$, in the sense the SMM estimator is more precise using $Z_{1}$
than when using $Z_{2}$ as an instrument. For example, in the original
design where both instruments are valid, using only $Z_{1}$ as an instrument
resulted in the median of the $10\mbox{,}000$ estimates to be equal to $0.6009$
with the interquartile range equal to $0.1967$, whereas using only $Z_{1}$
as an instrument resulted in a median of $0.6242$, with a much larger
interquartile range of $1.5253$. If the bias is due to a violation
of the CMI assumption for $Z_{1}$, the estimator based on $Z_{2}$ does not
have enough precision to reject the null that both moment conditions are
valid as frequently as for when $Z_{2}$ is invalid, as the estimator based
on $Z_{1}$ is more precise and the test has more power.

%s6.2 #&#
\subsection{Logistic SMM}\label{sec6.2}

To investigate the performance of the GMM estimators for the logistic SMM, we
generate data from population $M_{2}$ satisfying the logistic SMM model and
its corresponding NEM and CMI identification restrictions. More
specifically, the data are generated from
\begin{eqnarray*}
E ( Y|X,Z_{1},Z_{2} ) &=&\expit \bigl\{
\beta_{0}+ ( \beta _{1}+\psi_{0} ) X+
\beta_{2}Z_{1}\\
&&\hspace*{22pt}{}+\beta_{3}Z_{2}+\beta
_{4}XZ_{1}+\beta_{5}XZ_{2}\bigr\},
\end{eqnarray*}
where the treatment effect is again $\psi_{0}=0.6$. Similarly to model
$%
M_{1}$, we set $P ( Z=1 ) =0.3$, $P ( Z=2 ) =0.2$,
 $P ( X=1|Z=z ) =p_{10}+0.15\times z$, $E (
Y_{0} ) =0.19$, $E ( Y ) =0.25$, $\beta_{1}=0.15$,
$\beta
_{4}=-0.6$ and $\beta_{5}=0.6$. The other parameters are such that CMI and
NEM hold: $\beta_{0}=-1.518$, $\beta_{2}=0.3183$, $\beta_{3}= -0.5202$,
and $p_{10}=0.4404$.

\begin{table}%[b]
\tabcolsep=2pt
\caption{Monte Carlo estimation results for logistic SMM}\label{tab2}
\begin{tabular*}{\columnwidth}{@{\extracolsep{\fill}}ld{2.5}d{2.5}d{2.5}@{}}
\hline
 & \multicolumn{1}{c}{\textbf{Single instrument}} &
 \multicolumn{2}{c@{}}{\textbf{Multiple instruments}} \\
\hline
\textbf{Instruments} $\mathbf{S}$  & \multicolumn{1}{c}{$\bolds{1,Z}$} &
\multicolumn{1}{c}{$\bolds{1,Z_{1},Z_{2}}$}
 & \multicolumn{1}{c@{}}{$\bolds{1,Z_{1},Z_{2}}$} \\
 \hline
\textbf{Moment conditions}  & \multicolumn{1}{c}{\textbf{Joint/2SGMM}} &
\multicolumn{1}{c}{\textbf{2SGMM}} & \multicolumn{1}{c@{}}{\textbf{Joint-GMM}} \\
\hline
One-step GMM  &  & & \\
$\alpha_{0}$  & 0.1912 & 0.1905 & 0.1907 \\
 & (0.0168) & (0.0153) & (0.0153) \\
 & [0.0167] & [0.0152] & [0.0152] \\
$\psi_{0}$ & 0.5970 & 0.6033 & 0.6001 \\
   & (0.1905) & (0.1729) &  (0.1731) \\
   & [0.1899] & [0.1722] & [0.1721] \\
Two-step GMM  & & & \\
$\alpha_{0}$  & & 0.1904 & 0.1911 \\
   & & (0.0153) & (0.0154) \\
   & & [0.0152] & [0.0152] \\
$\psi_{0}$  & & 0.6038 & 0.5957 \\
   & & (0.1729) & (0.1735) \\
   & & [0.1722] & [0.1722] \\
Hansen \textit{J}  & & 0.9882 & 0.9827 \\
Rej. freq. 5\%  & & 0.0503 & 0.0495 \\
\hline
\end{tabular*}
\tabnotetext[]{tz}{Notes: Sample size 10,000; means based on 10,000
Monte Carlo replications;  std. [error] in brackets; means of estimated
standard errors in square brackets; data drawn from population model $M_2$
 as described in Section~\ref{sec6.2}; $\alpha_{0}=0.19$ and $\psi_{0}=0.6$.}
\end{table}

Table~\ref{tab2} contains estimation results for $10\mbox{,}000$ samples of size $10\mbox{,}000$
drawn from population model $M_{2}$. Three different versions of the GMM
estimator for the logistic SMM are applied: the first column of Table~\ref{tab2}
contains the results of the just-identified model using multivalued $Z$
as a
single instrument; in the second column, we present the one- and two-step
GMM estimates for the 2SGMM using multiple instruments; and the third column
contains the corresponding results for the joint-GMM estimator based on
(\ref{lmomjoint}). Both the 2SGMM and joint-GMM estimators use saturated
logistic models for $\bolds{\beta}$ as in (\ref{SatLog}).

All of the estimators are virtually unbiased and the means of the estimated
standard errors are close to Monte Carlo standard errors. There is an
efficiency gain from using the instruments separately: the standard error
in the just-identified case is 0.1905, compared to 0.1729 for the 2SGMM
estimator. The performances of the 2SGMM estimator and the GMM estimator
using the joint moment conditions are virtually identical. The Hansen
\textit{J}-tests are well behaved in both cases. There is no
efficiency gain
from using the two-step GMM estimators as compared to the one-step
estimators in this design.

As with the multiplicative SMM, we also find that the estimators behave well
for instruments with 6 or even 11 values, although we find that the
2SGMM
estimator has a small upward finite sample bias in the designs we
considered. For example, for an instrument with values $0,1,2,\ldots,10$, we
get means (std. error) of the two-step GMM estimates of 0.6323 (0.1073) for
2SGMM and 0.5999 (0.1066) for the joint moments GMM estimator. Details of
this design are available from the authors.

Finally, we return to the design with $Z$ taking the values $0,1,2$, and
modify population model $M_{2}$ so as to study how these estimators perform
when $Z$ is not a valid instrumental variable. We keep all parameters the
same but make the ``instrument'' $Z_{2}$
invalid, by changing the parameter of $Z_{2}$ to $\beta_{3}+\tau$
with $%
\tau=0.25$. The GMM estimators are now severely biased upwards. The
mean of
10,000 Monte Carlo estimates of the two-step GMM estimator using the joint
moments (\ref{lmomjoint}) is equal to 1.2805, with a standard error of
0.1511. However, in this case the mean (variance) of Hansen's \textit
{J}-test is equal to 1.26 (3.09), with a rejection frequency at the 5\% level
of only 8.5\%. In contrast, if we instead change the parameter of
$Z_{1}$ to
$\beta_{2}+\tau$ with $\tau=0.1$, the estimator has a much smaller bias,
with a mean of 0.5527 and standard error of 0.1660, but the \textit{J}-test
has much more power in this case as it rejects 49.4\% of the time at
the 5\%
level. This is explained by the fact that here $Z_{2}$ is a
stronger instrument than $Z_{1}$.

%s7 #&#
\section{Local Average Treatment Effects}\label{sec7}

The parameters of the SMMs we have considered thus far are all
identified by
the assumption of no effect modification by the instruments (NEM). For the
case where we have two instruments $Z_{1}$ and $Z_{2}$, recall that the NEM
assumption for the identification of the conditional causal relative
risk is
that
\[
\frac{E(Y|X,Z_{1},Z_{2})}{E(Y_{0}|X,Z_{1},Z_{2})}=\exp ( \psi _{0}X ) ,
\]
that is, the instruments $Z_{1}$ and $Z_{2}$ do not modify the causal
effect of
$X$ on the risk. In this section, we consider how the failure of NEM
impacts on GMM estimators for additive and multiplicative SMMs with
multiple instruments.

\citet{ClaWin10} review identification results concerning the
additive and multiplicative SMMs in the simple case of a single binary
instrument where both $X$ and $Y$ are also binary. If the NEM assumption
fails, then a causal effect is identified if the instrument $Z$ has causal
effect on treatment $X$ and selection is ``monotonic''. In this simple case,
where $Z$ is randomised treatment assignment and $X$ is the selected
treatment, selection is monotonic if
\[
P(X_{1}-X_{0}\geq0)=1,
\]
that is, subjects cannot defy their treatment assignments in every potential
scenario, so that $\{X_{1}=0,\allowbreak X_{0}=1\}$ has zero probability. Under
monotonicity, the additive SMM estimator (\ref{ClassicalIV})
identifies the
``local average treatment effect'' ($\mathrm{LATE}$), and the multiplicative
SMM identifies the ``local risk ratio'' ($\mathrm{LRR}$), where
\begin{eqnarray*}
\mathrm{LATE}&=&E(Y_{1}-Y_{0}|X_{1}>X_{0});\\
\mathrm{LRR}&=&\frac{%
E(Y_{1}|X_{1}>X_{0})}{E(Y_{0}|X_{1}>X_{0})}.
\end{eqnarray*}
$\mathrm{LATE}$ is the average treatment effect for the subgroup of subjects
who actually and counterfactually accept the treatments to which they have
been assigned, that is, $X_{1}=1$ and $X_{0}=0$; for this reason, these
subjects are also known as ``compliers'' and $\mathrm{LATE}$ is also known as
the ``complier average causal effect'' (CACE). The logistic SMM does not
estimate a local causal effect when NEM fails, but for binary outcomes the
local odds ratio can be estimated by taking the ratio of $\mathrm{LRR}$
estimates obtained by fitting multiplicative SMMs to binary $Y$ and $1-Y$.

If we have two instruments, then these instruments could in principle define
two different local causal effects, provided that the two instruments
can be
combined into a single multivalued instrument. We consider using the
single $%
K$-valued instrument $Z\in \{ 0,1,2,\ldots,K-1 \} $ for binary $X$.
In this scenario, monotonic selection does not have the convenient ``no
defiers'' interpretation; instead, selection is monotonic if
$z>\widetilde{z}$
implies that $X_{z}\geq X_{\widetilde{z}}$ with probability $1$, for
any two
values $z\neq\widetilde{z}$ of the instrument. From this, we can
define the
analogue of (\ref{ClassicalIV}) for $z>\widetilde{z}$ as
\[
\beta_{z,\widetilde{z}}=\frac{E(Y|Z=z)-E(Y|Z=\widetilde
{z})}{E(X|Z=z)-E(X|Z=%
\widetilde{z})},
\]
where $\beta_{z,\widetilde{z}}=E(Y_{1}-Y_{0}|X_{z}>X_{\widetilde
{z}})\equiv
\mathrm{LATE}_{z,\widetilde{z}}$ under monotonicity.

The 2SLS estimator for the additive SMM is obtained as the OLS estimator
from the regression of $Y$ on $\widehat{X}$, where $\widehat{X}$ is the
prediction from the first-stage regression of $X$ on $\mathbf{S}=
\{
1,Z_{1},\ldots,Z_{K-1} \} ^{\prime}$ and $Z_{k}=I(Z=k)$. Let monotonicity
hold and the values of $Z$ be ordered such that $E ( X|Z=k )
>E ( X|Z=k-1 ) $. \citet{ImbAng94} show that the 2SLS
estimator is consistent for
\[
\beta_{z}=\sum_{k=1}^{K-1}
\mu_{k}\beta_{k,k-1},
\]
where
\begin{eqnarray*}
\mu_{k}&=&\bigl\{E(X|Z=k)-E(X|Z=k-1)\bigr\}\\
&&{}\cdot\frac{\sum_{l=k}^{K-1}\{
E(X|Z=l)-E(X)\}\pi
_{l}}{\sum_{l=0}^{K-1}E(X|Z=l)\{E(X|Z=l)-E(X)\}\pi_{l}},
\end{eqnarray*}
and $\pi_{l}=P ( Z=l ) $ such that $0\leq\mu_{k}\leq1$
and $%
\sum_{l=1}^{K-1}\mu_{k}=1$; see also \citet{AngImb95} and Angrist and Pischke (2009). In other words, when NEM fails but selection is
monotonic, the 2SLS estimator is not consistent for $E(Y_{1}-Y_{0}|X=1)$,
but for a weighted sum of local average treatment effects.

Alternatively, if we define
\[
\beta_{k,0}=\frac{E ( Y|Z=k ) -E ( Y|Z=0 )
}{E (
X|Z=k ) -E ( X|Z=0 ) },
\]
then, following the proof given by \citet{AngImb95}, it is easily
established that
\[
\beta_{z}=\sum_{k=1}^{K-1}
\lambda_{k}\beta_{k,0},
\]
where
\begin{eqnarray*}
\lambda_{k}&=&\bigl\{E(X|Z=k)-E(X|Z=0)\bigr\}\\
&&{}\cdot\frac{\{E(X|Z=k)-E(X)\}\pi_{k}}{%
\sum_{l=0}^{K-1}E(X|Z=l)\{E(X|Z=l)-E(X)\}\pi_{l}},
\end{eqnarray*}
such that $\sum_{l=1}^{K-1}\lambda_{k}=1$. However, in this case,
$\beta
_{z}$ is only a weighted average of the $\beta_{k,0}$ (i.e., $0\leq
\lambda
_{k}\leq1$) if $E(X|Z=1)>E(X)$.

We now extend this result to the multiplicative SMM and give an analogous
result for local risk ratios. In Section~\ref{sec4.3} we established that the
one-step GMM estimator for $\exp ( -\psi_{0} ) $ using moment
condition (\ref{mmom0}) was equivalent to a linear 2SLS estimator
because
%
%e28 #&#
\begin{eqnarray}\label{linmult}
&&Y\exp ( -X\psi_{0} ) -\alpha_{0}\nonumber\\[-8pt]\\[-8pt]
&&\quad =Y ( 1-X ) +YX\exp (-
\psi_{0})-\alpha_{0}.\nonumber
\end{eqnarray}
We can therefore straightforwardly generalise the abo\-ve results of \citet{ImbAng94} for the additive SMM to the multiplicative SMM for the
inverse local risk ratio. As above, let
%
%e29 #&#
\begin{eqnarray}\label{Waldnonlin}
{\fontsize{9}{12}\selectfont{
 \begin{aligned}
& \qquad e_{k,k-1}^{-\beta}\\
&\qquad \quad =\frac{E\{Y(X-1)|Z=k\}-E\{Y ( X-1 )
|Z=k-1\}}{%
E ( YX|Z=k ) -E ( YX|Z=k-1 ) },
 \end{aligned}
}}
\end{eqnarray}\hspace*{-10pt}
where
\[
e_{k,k-1}^{-\beta}=\frac{E ( Y_{0}|X_{k}>X_{k-1} )
}{E (
Y_{1}|X_{k}>X_{k-1} ) }\equiv\mathrm{ILRR}
_{k,k-1}
\]
is the \textit{inverse} local risk ratio under monotonicity; see \citet{Ang01}. We then get equivalent results to the above for the linear SMM,
namely, the 2SLS estimator for $\exp ( -\psi_{0} ) $ in
(\ref%
{linmult}) is a consistent estimator of
\[
e_{z}^{-\beta}=\sum_{k=1}^{K-1}
\mu_{k}e_{k,k-1}^{-\beta},
\]
where
\begin{eqnarray*}
{\fontsize{10}{12}\selectfont{
 \begin{aligned}
\mu_{k}=&\bigl\{E(YX|Z=k)-E(YX|Z=k-1)\bigr\}\\
&{}\cdot\frac{\sum_{l=k}^{K-1}\{
E(YX|Z=l)-E(YX)\}%
\pi_{l}}{\sum_{l=0}^{K-1}E(YX|Z=l)\{E(YX|Z=l)-E(YX)\}\pi_{l}},
 \end{aligned}
}}
\end{eqnarray*}
and so $e_{z}^{-\beta}$ is a weighted average of inverse local risk ratios
if $E ( YX|Z=k ) >E ( YX|Z=k-1 ) $. As in \citet{AngImb95}, the weights $\mu_{k}$ are proportional to $%
E(YX|Z=k)-E(YX|Z=k-1)$, and hence the stronger the instrument, that is, the
bigger the impact of the instrument on the regressor $YX$ in~(\ref%
{linmult}), the more weight (\ref{Waldnonlin}) receives in the linear
combination. The second component of the weighting gives more weight to the
estimates (\ref{Waldnonlin}) when the values of $Z$ are closer to the center
of the distribution of $Z$ (see \cite*{AngImb95}, pages 437).

For the local risk ratio, we use the results from Section~\ref{sec4.3} that the
one-step GMM estimator for $\exp ( \psi_{0} ) $ can be obtained
from a linear IV estimator in the additive SMM with $YX$ as the ``outcome''
and $Y ( X-1 ) $ as the ``treatment'', but with instruments a
constant and $E(YX|S)$. Let
\begin{eqnarray*}
\hspace*{-4pt}&&e_{k,k-1}^{\beta}\\
\hspace*{-4pt}&&\quad =\frac{E ( YX|Z=k ) -E (
YX|Z=k-1 ) }{%
E\{Y(X-1)|Z=k\}-E\{Y ( X-1 ) |Z=k-1\}},
\end{eqnarray*}
where $e_{k,k-1}^{\beta
}=E(Y_{1}|X_{k}>X_{k-1})/E(Y_{0}|X_{k}>\break X_{k-1})\equiv\mathrm{LRR}_{k,k-1}$
under monotonicity. It follows that the multiplicative SMM estimator is
consistent for
\[
e_{z}^{\beta}=\sum_{k=1}^{K-1}
\tau_{k}e_{k,k-1}^{\beta},
\]
where
\begin{eqnarray*}
{\fontsize{9.5}{12}\selectfont{
  \begin{aligned}
\tau_{k} &=& \bigl\{ E \bigl( Y ( X-1 ) |Z=k \bigr) -E \bigl( Y ( X-1
) |Z=k-1 \bigr) \bigr\}
\\
&&{}\cdot\frac{\sum_{l=k}^{K-1}\{E(YX|Z=l)-E(YX)\}\pi_{l}}{%
\sum_{l=0}^{K-1}E\{Y ( X-1 ) |Z=l\}\{E(YX|Z=l)-E(YX)\}\pi_{l}},
 \end{aligned}
}}
\end{eqnarray*}
and hence $e_{z}^{\beta}$ is a weighted average of local risk ratios
if $%
E(YX|Z=k)>E(YX|Z=k-1)$ and $E \{ Y ( X-1 ) |Z=k \}
>E \{ Y ( X-1 ) |Z=k-1 \} $.

As an example, consider an instrument that takes the values $Z= \{
0,1,2,3 \} $, with $Y$ and $X$ generated from a bivariate normal
distribution as
\begin{eqnarray*}
X &=&I(c_{0}+c_{1}Z_{1}+c_{2}Z_{2}+c_{3}Z_{3}-V>0),
\\
Y &=&I(b_{0}+b_{1}X-U>0),
\\
\lleft( %
\begin{array} {c} U
\\
V%
\end{array} %
 \rright) &\thicksim&N\lleft( \pmatrix{0\cr 0} ,
\pmatrix{1 & \rho\cr \rho& 1}
  \rright) ,
\end{eqnarray*}
with, as before, $Z_{k}=I(Z=k)$. Setting $\pi_{l}=P ( Z=l ) =0.25$
for all $l$, the $c_{l}$ parameters are such that $P (
X=1|Z=l )
=0.1+0.1\times l$, $b_{0}=\Phi^{-1}(0.4)$, \mbox{$b_{1}=0.5$} and $\rho
=0.8$. The
local risk ratios in this population are $\mathrm{LRR}_{1,0}=1.1585$,
$\mathrm{%
LRR}_{2,1}=1.3227$ and $\mathrm{LRR}_{3,2}=1.5303$; the population $\tau
$-weights are
\[
\tau_{1}=0.3725,\quad  \tau_{2}=0.3991,\quad  \tau_{3}=0.2285.
\]
\citet{ClaWin10} show that the NEM assumption does not hold
under this design. However, the instruments are monotonic and so the
one-step GMM estimator based on moment conditions (\ref{mmom0}) identifies
the weighted average $\tau_{1}\mathrm{LRR}_{1,0}+\tau_{2}\mathrm{LRR}%
_{2,1}+\tau_{3}\mathrm{LRR}_{3,2}=1.3090$. Table~\ref{tab3} presents some estimation
results confirming this, for a sample of size 40,000 and for 10,000 Monte
Carlo replications. Using the two-step GMM results, the Hansen \textit
{J}-test rejects the null 47\% of the time at the 5\% level, therefore clearly
having power to reject this violation of the NEM assumption.

\begin{table}
\tabcolsep=0pt
\caption{Risk ratio estimation results}\label{tab3}
\begin{tabular*}{\columnwidth}{@{\extracolsep{\fill}}lccccccc@{}}
\hline
& $\bolds{e_{1,0}^{\beta}}$ & $\bolds{e_{2,1}^{\beta}}$
& $\bolds{e_{3,2}^{\beta}}$ &
$\bolds{e_{z}^{\beta}}$ & $\bolds{\tau_{1}}$ & $\bolds{\tau_{2}}$
& $\bolds{\tau_{3}}$  \\
\hline
Mean &1.1644 & 1.3304 &
1.5415 & 1.3113 &0.3726 & 0.3995 & 0.2279 \\
St. dev. & 0.0946 & 0.1213 &
0.1601 & 0.0377 &
0.0268 & 0.0321 & 0.0216  \\
\hline
\end{tabular*}
\tabnotetext[]{tz}{Notes: Estimation results from 10,000 Monte Carlo
replications. Sample size 40,000.}
\end{table}

%s8 #&#
\section{The Effect of Adiposity on Hypertension}\label{sec8}

%s8.1 #&#
\subsection{Binary Exposure}\label{sec8.1}

\citet{Timetal09} used multiple genetic instruments to estimate the
causal effect of adiposity on hypertension from the Copenhagen General
Population Study; full details of the variable definitions and selection
criteria are given in that paper. We apply the procedures described
above to
reanalyse these data using additive, multiplicative and logistic SMMs, using
the same genetic markers as instruments for adiposity. Furthermore, our
sample includes additional individuals who have been recruited into the
study since the previous study was published; the total number of
individuals in our analyses is 55,523.

The binary outcome variable is an indicator of whether an individual has
hypertension, which is defined as a systolic blood pressure of
$>$140 mmHg, diastolic blood pressure of $>$90 mmHg, or the taking
of antihypertensive drugs. The intermediate adiposity phenotype is being
overweight, defined as having a $\mathrm{BMI}>25$. The two Single
Nucleotide Polymorphisms (SNPs) that were used as instruments by
\citet{Timetal09} and that have been consistently shown to relate to BMI and
adiposity are the \textit{FTO} (rs9939609) and \textit{MC4R} (rs17782313)
loci; see \citet{Fraetal07} and \citet{Looetal08}. \citet{Lawetal08} provide further details on the use of genes as instruments in
Mendelian randomisation studies.

\textit{FTO} is specified as having three categories: no risk alleles
(homozygous TT), one risk allele (heterozygous AT) and two risk alleles
(homozygous AA). Due to the nature of the association between \textit{MC4R}
and adiposity (a dominant genetic model), \textit{MC4R} is specified as
having two categories: no risk alleles (TT) versus one or two risk alleles
(CT or CC). Combining the two instruments together results in an instrument
with 6 different values, but we found that two pairs of combinations of
alleles gave the same predicted value of being overweight; this is also true for the projection in the multiplicative SMM.
We therefore condensed the number
of values of the instrument to four. The combinations for the four values
are given in Table~\ref{tab4}. Table~\ref{tab5} gives the frequency distributions for the
hypertension $ ( Y ) $ and overweight $ ( X ) $ variables.

\begin{table}
\caption{Combinations of instruments}\label{tab4}
\begin{tabular*}{\columnwidth}{@{\extracolsep{\fill}}lccc@{}}
\hline
\textbf{FTO} & \textbf{MC4R} & $\bolds{Z}$ & \textbf{Freq.}  \\
\hline
0 & 0 & 0 & 0.20 \\
0 & 1 & 1 & 0.15 \\
1 & 0 & 1 & 0.27 \\
1 & 1 & 2 & 0.21 \\
2 & 0 & 2 & 0.09 \\
2 & 1 & 3 & 0.07 \\
\hline
\end{tabular*}
\end{table}

\begin{table}[b]
\tabcolsep=0pt
\caption{Frequency distributions for the hypertension $ ( Y ) $ and overweight $ ( X ) $
variables}\label{tab5}
\begin{tabular*}{\columnwidth}{@{\extracolsep{\fill}}lcccccccccc@{}}
\hline
& \multicolumn{2}{c}{\textbf{All}} & \multicolumn{2}{c}{$\bolds{Z=0}$} &
\multicolumn{2}{c}{$\bolds{Z=1}$} & \multicolumn{2}{c}{$\bolds{Z=2}$} &
\multicolumn{2}{c@{}}{$\bolds{Z=3}$} \\
\hline
& \multicolumn{2}{c}{$\bolds{X}$} & \multicolumn{2}{c}{$\bolds{X}$} & \multicolumn
{2}{c}{$\bolds{X}$} & \multicolumn{2}{c}{$\bolds{X}$} & \multicolumn{2}{c}{$\bolds{X}$} \\
\hline
$\bolds{Y}$ & $\mathbf{0}$ & $\mathbf{1}$ & $\mathbf{0}$ & $\mathbf{1}$ & $\mathbf{0}$
& $\mathbf{1}$ & $\mathbf{0}$ &
$\mathbf{1}$ & $%
\mathbf{0} $ & $\mathbf{1}$ \\
\hline
$0$ & 0.18 & 0.12 & 0.19 & 0.12 &  0.19 & 0.12 & 0.17 & 0.13 & 0.16 & 0.13 \\
$1$ & 0.25 & 0.44 & 0.27 & 0.42 & 0.26 & 0.43 & 0.23 & 0.46 & 0.23 & 0.48 \\
\hline
\end{tabular*}
\end{table}

\begin{table*}
\tabcolsep=0pt
\caption{SMM estimation results of the effect of being
overweight on hypertension}\label{tab6}
\begin{tabular*}{\textwidth}{@{\extracolsep{\fill}}lcccc@{}}
\hline
Additive & OLS & 2SLS & GMM2 & $J$-test \\
$\psi_{0}$ &0.2009 &0.2091&0.2095&0.2956\\
 &  $[ 0.1932;0.2087 ]$&$[ 0.0485;0.3697]$&$[ 0.0489;0.3701]$&\\ [6pt]
 Multiplicative & Gamma & GMM1 & GMM2 & $J$-test \\
 $\exp ( \psi_{0} ) $ &1.3464 &1.3621&1.3640&0.3071\\
   &$[ 1.3300;1.3630]$ &$[ 1.0784;1.7204]$&$[ 1.0798;1.7231]$&\\[6pt]
Logistic & Logistic regression & GMM1 & GMM2 & $J$-test \\
$\exp ( \psi_{0} ) $&2.5823 &2.8317 & 2.8656&0.2924 \\
& $[ 2.4885;2.6797]$&$[ 1.2382;6.4759]$ & $[ 1.2538;6.5489]$& \\
  \hline
\end{tabular*}
\tabnotetext[]{tz}{Notes: Sample size 55,523. Gamma regression uses
log link; multiplicative SMM uses moments (\ref{mmom0}); logistic SMM uses
joint moments (\ref{lmomjoint}); instruments, $ S= \{ 1, Z_{
1}, Z_2, Z_3 \} $; 95\% CIs in brackets; $p$-values are reported for
the $ J$-test.}
\end{table*}

The estimation results for the linear, multiplicative and logistic SMM
estimators are presented in Table~\ref{tab6}. The instrument set for the GMM
estimators is $\mathbf{S}=(1,Z_{1},Z_{2},Z_{3})^{\prime}$. For the linear
SMM, the 2SLS and two-step GMM estimates are virtually identical to the OLS
estimate. As the F-statistic in the regression of overweight on
$\mathbf{S}$
is equal to 113, this is not due to a weak instrument problem. The OLS
estimate of the risk difference is quite large and equal to 0.20 (95\% CI
0.19; 0.21). The two-step GMM estimate is almost the same and equal to 0.21
(95\% CI 0.05--0.37), but clearly the 95\% confidence interval is much wider
for the two-step GMM estimate than it is for OLS. The \textit{J}-test does
not reject the null of the validity of the model assumptions, including the
NEM assumption, and therefore these results indicate that there may not be
much confounding bias in the OLS results. We find similar results for the
multiplicative and logistic SMMs. The GMM estimates are virtually identical
to the Gamma and the logistic regression estimates, respectively, and all
estimates indicate that being overweight leads to hypertension. The Gamma
estimate for the risk ratio is equal to 1.35 (95\% CI, 1.33--1.36), whereas
the two-step GMM estimate is equal to 1.36 (95\% CI 1.08--1.72). We present
and compare the multiplicative SMM results to that of the Gamma generalised
linear model with a log link here, because moment conditions (\ref
{mmom0})--(%
\ref{mmomc}) when using $X$ as an instrument for itself are equivalent to
the first-order condition of the Gamma with log link GLM. The logistic
regression odds ratio is equal to 2.58 (95\% CI, 2.49--2.68) and the two-step
GMM estimate is equal to 2.87 (95\% CI 1.25--6.55). All estimation results
indicate a large causal effect of adiposity on hypertension.

%s8.2 #&#
\subsection{Continuous Exposure}\label{sec8.2}

Following \citet{VanGoe03}, we can use the same GMM
format to estimate the logistic SMM with a continuous exposure $X$.
With a
continuous exposure, parametric modelling assumptions have to be made in
order to identify causal parameters. As in \citet{VanGoe03} and \citet{Vanetal11}, we impose that the exposure
effect is
linear in the exposure on the log-odds ratio scale and independent of the
instrumental variable:
\[
\frac{\operatorname{odds} ( Y=1|X,Z ) }{\operatorname{odds}(Y_{0}=1|X,Z)}=\exp ( \xi_{0}X ) ,
\]
where $\operatorname{odds} ( Y=1|X,Z ) =P(Y=1|X,Z)/P(Y=0|X,Z)$.
Further, we
specify the association model as
\begin{eqnarray*}
\logit  \bigl\{ P(Y=1|X,Z) \bigr\} &=&\logit  \bigl\{
m_{\beta
}(X,Z_{1},Z_{2},Z_{3}) \bigr\}
\\
&=&\beta_{0}+\beta_{1}X+\beta_{2}Z_{1}+
\beta_{3}Z_{2}\\
&&{}+\beta _{4}Z_{3}+
\beta_{5}XZ_{1}+\beta_{6}XZ_{2}\\
&&{}+
\beta_{7}XZ_{3},
\end{eqnarray*}
and estimate the parameters using the joint moment conditions as in
(\ref%
{lmomjoint}).

For the continuous exposure we use $ ( \mathit{BMI}-\overline{\mathit{BMI}} )
$, $%
10 ( \ln \mathit{BMI}-\overline{\ln \mathit{BMI}} ) $ and $10 ( \ln
\mathit{RELBMI} )
$, where   $\ln \mathit{BMI}$ is the natural logarithm of $\mathit{BMI}$, and
$\ln
\mathit{RELBMI}$ are the residuals of the regression of $\ln \mathit{BMI}$ on sex, age, age
squared, ln(height) and an age--sex interaction, as used in \citet{Timetal09} to represent relative BMI. We subtract the mean from $\mathit{BMI}$ and
$\ln
\mathit{BMI}$ to ensure that zero exposure is part of the data range. We further
multiply the $\ln \mathit{BMI}$ and $\ln \mathit{RELBMI}$ by a factor $10$ so that the
estimated odds ratio is for an increase in exposure of approximately
$10\%$.

Table~\ref{tab7} presents the two-step estimation results for three separate models
for the three exposure measures. Again, we find a strong positive
effect of
adiposity on hypertension. The estimate of the odds ratio for a one-unit
increase in BMI is equal to 1.12 (95\% CI 1.10; 1.67), whereas the estimates
for the odds ratios for a 10\% increase in $\ln \mathit{BMI}$ or $\ln \mathit{RELBMI}$ are
1.35 (95\% CI 1.10--1.67) and 1.33 (95\% CI 1.09--1.63), respectively, the
latter two therefore virtually identical. Also, for these logistic SMM models
with continuous exposures, the $J$-test results do not indicate a problem
with the model assumptions.

\begin{table}[t]
\tabcolsep=0pt
\caption{Estimation results for double-logistic SMM with
continuous exposure}\label{tab7}
\begin{tabular*}{\columnwidth}{@{\extracolsep{\fill}}lccc@{}}
\hline
Exposure & $\mathit{BMI}$ & $\ln \mathit{BMI}$ & $\ln \mathit{RELBMI}$  \\ [6pt]
$\exp ( \xi_{0} ) $ &1.1187  &1.3546&1.3337\\
&  $[ 1.0984;1.6705 ]$& $[1.0984;1.6705]$& $[1.0929;1.6276]$\\
[6pt]
\textit{J}-test & 0.4714 & 0.4828 & 0.5004  \\
\hline
\end{tabular*}
\tabnotetext[]{tz}{Notes: Sample size 55,523. Two-step GMM estimates,
using joint moments (\ref{lmomjoint}). Instruments, $\mathbf{S}= \{ 1, Z_1,
Z_2, Z_3 \} $. $ \mathit{BMI}$ and ${\ln\mathit{BMI}}$ taken in deviation from the mean.
$\ln \mathit{BMI}$
 and $\ln \mathit{RELBMI}$ multiplied by a factor 10. 95\% CIs in brackets;
 $p$-values are reported for
the \textit{J}-test.}
\end{table}

%s9 #&#
\section{Discussion}\label{sec9}

We have shown how the conditional moment conditions that identify additive,
multiplicative and logistic SMMs can be used to derive a standard GMM
estimator of the type widely used in econometrics. The key to this
formulation is simply to treat the expected exposure-free potential
outcome $%
E ( Y_{0} ) $ as a parameter. For simple SMMs without continuous
baseline covariates, these estimators are semiparametrically efficient if
the identifying instrumental variables are orthogonal binary variables. In
these cases, the estimator combines the instruments optimally in the manner
proposed by \citet{BowVan11}. Another major advantage is that
standard GMM routines are available in statistical software packages. We
provide example Stata and R syntax in the \hyperref[app]{Appendix} for use by applied
researchers. These estimation routines provide correct asymptotic inference,
even for the logistic SMM, when the two sets of model parameters are
estimated jointly, and a simple test for the validity of the SMM moment
conditions. We used Monte Carlo studies to show that the Hansen \textit{J}-test can have power to detect violations of the CMI and NEM assumptions.
Moreover, if the NEM assumption fails and selection is monotonic, then we
have shown that the one-step GMM estimator for the multiplicative SMM is
consistent for a weighted average of the instrument-specific local risk
ratios.

A characteristic of all estimating equations for  SMMs is that the analyst
must specify and estimate auxiliary models further to the SMM.
Extending the
discussion in Section~\ref{sec2.3} to multiple instrumental variables, the estimating
equations for G-estimation depend on $E(Z_{j})=\mu_{j}$, which must be
replaced in the estimating equation by a consistent estimator $\widehat
{\mu
}_{j}$. To derive the correct asymptotic distribution, the moment conditions
for $\widehat{\mu}_{j}$ must be included in the system of moment
conditions. For the multiplicative SMM with multiple instruments discussed
in Section~\ref{sec4}, the extended set of moment conditions is
%
%e30 #&#
\begin{equation}
\lleft( %
\begin{array} {c} E(Z_{1}-\mu_{1})
\\
E(Z_{2}-\mu_{2})
\\
E \bigl\{ ( Z_{1}-\mu_{1} ) Y\exp ( -\psi _{0}X
) \bigr\}
\\
E \bigl\{ ( Z_{2}-\mu_{2} ) Y\exp ( -\psi _{0}X
) \bigr\}%
\end{array} %
 \rright) =\lleft( %
\begin{array} {c} 0
\\
0
\\
0
\\
0%
\end{array} %
 \rright) . \label{SMMEXP}
\end{equation}
The extended moment conditions can easily be incorporated in the Stata
and R
GMM estimation routines, and we include in the \hyperref[app]{Appendix} code that does this
for the additive, multiplicative and logistic SMMs.

There are two relative weaknesses of our approach in applications where
covariates $\mathbf{C}$ are required for identification, in other words,
where CMI only holds covariate conditionally such that
$E(Y_{0}|Z,\mathbf{C}%
)=E(Y_{0}|\mathbf{C})$ but $E(Y_{0}|Z)\neq E(Y_{0})$. To discuss these
weaknesses, consider a multiplicative SMM which does not depend on
$\mathbf{C%
}$ but where covariates are still required for identification. In terms
of a
GMM estimator, the unconditional moment conditions [equivalent to (\ref
{mmom0}) in Section~\ref{sec4.2}] are
%
%e31 #&#
\begin{equation}
E\lleft[ \biggl\{ \frac{Y}{\exp ( \psi_{0}X )
}-E(Y_{0}|
\mathbf{C}%
) \biggr\} \lleft( %
\begin{array} {c} \mathbf{S}
\\
\mathbf{C}%
\end{array} %
 \rright) \rright] =
\mathbf{0}, \label{multey0c}
\end{equation}
which can be seen to depend on the extended instrument $(\mathbf
{S}^{\prime
},\mathbf{C}^{\prime})^{\prime}$ and $E(Y_{0}|\mathbf{C)}$ as well
as the
SMM itself.

The first weakness is that the efficiency result for two-step GMM discussed
above does not hold if $\mathbf{C}$ includes continuous covariates or if
the resulting\vadjust{\goodbreak} extended instrument cannot otherwise be represented by a set
of the mutually orthogonal binary variables. In such scenarios, the two-step
GMM estimator is only locally efficient given the unconditional moments,
which here are (\ref{mmom0}). \citet{New93} discusses different
approaches to
improve efficiency, for example, using a power-series expansions of the
instruments.

The second weakness is that consistency of the GMM estimator now
depends on
the model for  $E(Y_{0}|\mathbf{C)}$ being correctly specified. By
definition, this model cannot be empirically tested for misspecification
because it is determined by the SMM; but the consequence of misspecifying
it is an inconsistent GMM estimator. In contrast, the G-estimators and the
double-logistic SMM estimator discussed in Section~\ref{sec2} require only that
$E(Z|%
\mathbf{C)}$ is correctly specified, which can be empirically tested for
misspecification. Likewise, the doubly robust estimating equations proposed
by \citet{Tan10} depend on covariate-conditional models for $Z$, $X$ given $Z$,
and $Y$ given $X$ and $Z$, all of which can be tested for misspecification.
The doubly robust property is attractive in theory, but these
estimators are
not available in standard software, and further work is required to explore
fully, rather than locally, efficient choices of weights for the estimating
equations. Further work on the GMM estimators proposed here with continuous
covariates might investigate the bias and efficiency of GMM estimators,
both asympotically and in finite samples, compared to existing estimators
for SMMs; see \citet{Okuetal12}.

\begin{appendix}

\section*{Appendix: Stata and R Syntax}\label{app}

In this section we present example Stata (version 11) and R (version 2.13.1)
syntax to fit SMMs using generalised method of moments routines. Our example
code uses the notation of $Y$ the outcome, $X$ the exposure and two
instrumental variables, $Z_{1}$, $Z_{2}$, in addition to the constant vector
of 1's. Both syntaxes easily generalise to more instruments and allow
different association models in the double logistic SMM.

In both Stata and R it is possible to specify analytic first derivatives,
which we find greatly reduces the time for the models to fit. Also, both
syntaxes allow the inclusion of covariates. We have not included these extra
syntaxes here but they are available on request.

\subsection*{Stata Syntax}

The Stata syntax uses the \texttt{gmm} command; and \texttt{\{ey0\}} denotes
$E ( Y_{0} ) $ the mean exposure free potential outcome. After
fitting each SMM using two-step estimation we perform the Hansen
over-identification test using the \texttt{estat overid} post-estimation
command. The \texttt{gmm} command automatically includes a vector of
1's as
instruments to allow estimation of the constant [$E ( Y_{0}
) $]
term, hence, we just need to list \texttt{z1} and \texttt{z2} in the
\texttt{%
instruments()} option.

\subsubsection*{Additive SMM}

Here \texttt{\{psi\}} denotes the causal effect (which is a risk difference
for a binary outcome).

{\footnotesize
%TCIMACRO{%
%gmm (y - {ey0} - x*{psi}), instruments(z1 z2)
%estat overid
%BeginExpansion
%
\vspace*{5pt}\begin{verbatim}
gmm (y - {ey0} - x*{psi}),
     instruments(z1 z2)
estat overid
\end{verbatim}\vspace*{5pt}
%
%EndExpansion
}

%TCIMACRO{\TeXButton{noindent}{ }}%
%BeginExpansion
%
%EndExpansion
This is equivalent to Stata's built in \texttt{ivregress} command.

{\footnotesize
%TCIMACRO{%
%ivregress gmm y (x = z1 z2)
%estat overid
%BeginExpansion
%
\vspace*{5pt}\begin{verbatim}
ivregress gmm y (x = z1 z2)
estat overid
\end{verbatim}
%
%EndExpansion
}

\subsubsection*{Multiplicative SMM}

Here \texttt{\{psi\}} denotes the log causal risk ratio, and hence we
display the exponentiated estimate using the \texttt{lincom} command with
its \texttt{eform} option after fitting the model.

{\footnotesize
%TCIMACRO{%
%gmm (y*exp(-1*x*{psi}) - {ey0}), instruments(z1 z2)
%lincom [psi]_cons, eform // causal risk ratio
%estat overid
%BeginExpansion
%
\vspace*{5pt}\begin{verbatim}
gmm (y*exp(-1*x*{psi}) - {ey0}),
     instruments(z1 z2)
lincom [psi]_cons, eform //
causal risk ratio
estat overid
\end{verbatim}
%
%EndExpansion
}

We also give the Stata syntax for the alternative Multiplicative SMM
moments. Here \texttt{\{logey0\}} denotes $\log \{ E (
Y_{0} )
 \} $ and so we additionally display the exponentiated form of this
parameter after fitting the model.

{\footnotesize
%TCIMACRO{%
%gmm (y*exp(-x*{psi} - {logey0}) - 1), instruments(z1 z2)
%lincom [psi]_cons, eform // causal risk ratio
%lincom [logey0]_cons, eform // E[Y(0)]
%estat overid
%BeginExpansion
%
\vspace*{5pt}\begin{verbatim}
gmm (y*exp(-x*{psi} - {logey0}) - 1),
     instruments(z1 z2)
lincom [psi]_cons, eform //
causal risk ratio
lincom [logey0]_cons, eform // E[Y(0)]
estat overid
\end{verbatim}\vspace*{5pt}
%
%EndExpansion
}

%pa9.subsection.subsubsection.1 #&#
\paragraph*{Expanded moments for multiplicative SMM}
\mbox{}\vspace*{5pt}

{\footnotesize
%TCIMACRO{%
%gmm (z1-{mu1}) ///
%(z2-{mu2}) ///
%((z1-{mu1})*(y*exp(-1*x*{psi}))) ///
%((z2-{mu2})*(y*exp(-1*x*{psi}))) , ///
%winitial(identity)
%lincom [psi]_cons, eform // causal risk ratio
%estat overid
%BeginExpansion
%
\begin{verbatim}
gmm (z1-{mu1}) ///
(z2-{mu2}) ///
((z1-{mu1})*(y*exp(-1*x*{psi}))) ///
((z2-{mu2})*(y*exp(-1*x*{psi}))) , ///
winitial(identity)
lincom [psi]_cons, eform //
causal risk ratio
estat overid
\end{verbatim}
%
%EndExpansion
}

\subsubsection*{Logistic SMM}

Here \texttt{\{psi\}} denotes the log causal odds ratio. In the joint
estimation we use the \texttt{gmm} command's linear predictor substitution
syntax (we denote the linear predictor for the association model by
\texttt{\{xb:\}}). We collect the association and causal model parameter estimates
in a matrix called \texttt{from}; we then use these estimates as initial
values in the joint estimation. Also, in the joint estimation we
specify the
\texttt{winitial(unadjusted, independent)} option so that the moments are
assumed to be independent in the first step of estimation. Note in
Stata, $%
\operatorname{invlogit}(x)=\expit (x)=e^{x}/(1+e^{x})$.

{\footnotesize
%TCIMACRO{%
%* generate interactions
%gen xz1 = x*z1
%gen xz2 = x*z2
%
%* association model
%logit y x z1 z2 xz1 xz2
%matrix from = e(b)
%predict xblog, xb
%
%* causal model with incorrect SEs
%gmm (invlogit(xblog - x*{psi}) - {ey0}), instruments(z1 z2)
%matrix from = (from,e(b))
%
%* joint estimation of association and causal models
%gmm (y - invlogit({xb:x z1 z2 xz1 xz2} + {b0})) ///
%(invlogit({xb:} + {b0} - x*{psi}) - {ey0}), ///
%instruments(1:x z1 z2 xz1 xz2) ///
%instruments(2:z1 z2) ///
%winitial(unadjusted, independent) from(from)
%lincom [psi]_cons, eform // causal odds ratio
%estat overid
%}}%
%BeginExpansion
%
\vspace*{5pt}\begin{verbatim}
* generate interactions
gen xz1 = x*z1
gen xz2 = x*z2

* association model
logit y x z1 z2 xz1 xz2
matrix from = e(b)
predict xblog, xb

* causal model with incorrect SEs
gmm (invlogit(xblog - x*{psi}) - {ey0}),
     instruments(z1 z2)
matrix from = (from,e(b))

* joint estimation of association and
causal models
gmm (y - invlogit({xb:x z1 z2 xz1 xz2}
 + {b0})) ///
(invlogit({xb:} + {b0} - x*{psi})
- {ey0}), ///
instruments(1:x z1 z2 xz1 xz2) ///
instruments(2:z1 z2) ///
winitial(unadjusted, independent) from(from)
lincom [psi]_cons, eform //
causal odds ratio
estat overid
\end{verbatim}
%
%EndExpansion
}

\subsection*{R syntax}

The R syntax uses the \texttt{gmm()} function in the GMM package
(\citet{Cha10}), which we first load using \texttt{library(gmm)}. After
fitting each
SMM using two-step estimation we perform the Hansen over-identification test
using the \texttt{specTest()} function. The R code assumes our data is
in a
matrix called \texttt{data} whose columns contain the values of the
variables $Y$, $X$, $Z_1$ and $Z_2$ in this order with column names
\texttt{%
"y"}, \texttt{"x"}, \texttt{"z1"}, \texttt{"z2"}.

In this code we have specified the \texttt{vcov="iid"} option which assumes
the moment conditions are independent. We find specifying this option is
necessary for the models to converge on reasonably sized data sets. We also
find that changing the optimization algorithm used in the estimation through
the \texttt{method} option can reduce the time it takes the models to fit
(we find the BFGS and L-BFGS-B methods are the fastest).

\subsubsection*{Additive SMM}

First, we fit the Additive SMM using the \texttt{gmm()} function's formula
syntax for linear models.

{\footnotesize
%TCIMACRO{%
%asmm <- gmm(data[,"y"] ~ data[,"x"], x=data[,c("z1","z2")], vcov="iid")
%print(summary(asmm))
%print(cbind(coef(asmm),confint(asmm))) # estimates
%print(specTest(asmm))
%BeginExpansion
%
\vspace*{5pt}\begin{verbatim}
asmm <- gmm(data[,"y"] ~ data[,"x"],
x=data[,c("z1","z2")], vcov="iid")
print(summary(asmm))
print(cbind(coef(asmm),confint(asmm)))
# estimates
print(specTest(asmm))
\end{verbatim}\vspace*{5pt}
%
%EndExpansion
}

We can also pass the moment conditions to \texttt{gmm()} using its function
syntax. In order to do this, we first define a function \texttt{asmmMoments()}
which returns the ASMM moments. This function must have two arguments; the
first of which \texttt{theta} denotes the vector of parameters to be
estimated, where \texttt{theta[1]} is $E ( Y_{0} ) $ and
\texttt{%
theta[2]} is the causal risk difference. The second argument \texttt
{x} is
the data matrix; the user must avoid confusion here with the single variable
\texttt{X}. In the \texttt{gmm()} function the \texttt{t0} option specifies
the initial values of the parameter estimates. After we have fitted the
model with the call to \texttt{gmm()} we print out the model summary, then
the estimates and their 95\% CIs, and finally the over-identification test
using \texttt{specTest()}.

{\footnotesize
%TCIMACRO{%
%asmmMoments <- function(theta,x){
%# extract variables from x
%Y <- x[,"y"]
%X <- x[,"x"]
%Z1 <- x[,"z1"]
%Z2 <- x[,"z2"]
%# moments
%m1 <- (Y - theta[1] - theta[2]*X)
%m2 <- (Y - theta[1] - theta[2]*X)*Z1
%m3 <- (Y - theta[1] - theta[2]*X)*Z2
%return(cbind(m1,m2,m3))
%}
%
%asmm2 <- gmm(asmmMoments, x=data, t0=c(0,0), vcov="iid")
%print(summary(asmm2))
%print(cbind(coef(asmm2),confint(asmm2))) # estimates
%print(specTest(asmm2))
%BeginExpansion
%
\vspace*{5pt}\begin{verbatim}
asmmMoments <- function(theta,x){
# extract variables from x
Y <- x[,"y"]
X <- x[,"x"]
Z1 <- x[,"z1"]
Z2 <- x[,"z2"]
# moments
m1 <- (Y - theta[1] - theta[2]*X)
m2 <- (Y - theta[1] - theta[2]*X)*Z1
m3 <- (Y - theta[1] - theta[2]*X)*Z2
return(cbind(m1,m2,m3))
}

asmm2 <- gmm(asmmMoments, x=data, t0=c(0,0),
vcov="iid")
print(summary(asmm2))
print(cbind(coef(asmm2),confint(asmm2)))
# estimates
print(specTest(asmm2))
\end{verbatim}
%
%EndExpansion
}

\subsubsection*{Multiplicative SMM}

We again use the \texttt{gmm()} function syntax to fit the Multiplicative
SMM. First we define the function \texttt{msmmMoments()} to return the
\mbox{moments}. After fitting the model we print the model summary. Here
\texttt{%
theta[2]} is the log causal risk ratio, and so we print the exponentiated
form of this parameter.

{\footnotesize
%TCIMACRO{%
%msmmMoments <- function(theta,x){
%# extract variables from x
%Y <- x[,"y"]
%X <- x[,"x"]
%Z1 <- x[,"z1"]
%Z2 <- x[,"z2"]
%# moments
%m1 <- (Y*exp(- X*theta[2]) - theta[1])
%m2 <- (Y*exp(- X*theta[2]) - theta[1])*Z1
%m3 <- (Y*exp(- X*theta[2]) - theta[1])*Z2
%return(cbind(m1,m2,m3))
%}
%
%msmm <- gmm(msmmMoments, x=data, t0=c(0,0), vcov="iid")
%print(summary(msmm))
%print(exp(cbind(coef(msmm), confint(msmm))[2,])) # causal risk ratio
%print(cbind(coef(msmm), confint(msmm))[1,]) # E[Y(0)]
%print(specTest(msmm))
%BeginExpansion
%
\vspace*{5pt}\begin{verbatim}
msmmMoments <- function(theta,x){
# extract variables from x
Y <- x[,"y"]
X <- x[,"x"]
Z1 <- x[,"z1"]
Z2 <- x[,"z2"]
# moments
m1 <- (Y*exp(- X*theta[2]) - theta[1])
m2 <- (Y*exp(- X*theta[2]) - theta[1])*Z1
m3 <- (Y*exp(- X*theta[2]) - theta[1])*Z2
return(cbind(m1,m2,m3))
}

msmm <- gmm(msmmMoments, x=data, t0=c(0,0),
vcov="iid")
print(summary(msmm))
print(exp(cbind(coef(msmm),
confint(msmm))[2,])) # causal risk ratio
print(cbind(coef(msmm), confint(msmm))[1,])
# E[Y(0)]
print(specTest(msmm))
\end{verbatim}\vspace*{5pt}
%
%EndExpansion
}

We can also fit the alternative MSMM moments in the same way. Here
\texttt{%
theta[1]} denotes $\log \{ E ( Y_{0} )  \}$,
and so we
print out the exponentiated form of both estimates:

{\footnotesize
%TCIMACRO{%
%msmmAltMoments <- function(theta,x){
%# extract variables from x
%Y <- x[,"y"]
%X <- x[,"x"]
%Z1 <- x[,"z1"]
%Z2 <- x[,"z2"]
%# moments
%m1 <- (Y*exp(-theta[1] - X*theta[2]) - 1)
%m2 <- (Y*exp(-theta[1] - X*theta[2]) - 1)*Z1
%m3 <- (Y*exp(-theta[1] - X*theta[2]) - 1)*Z2
%return(cbind(m1,m2,m3))
%}
%
%msmm2 <- gmm(msmmAltMoments, x=data, t0=c(0,0), vcov="iid")
%print(exp(cbind(coef(msmm2), confint(msmm2)))) # exponentiate estimates
%print(specTest(msmm2))
%BeginExpansion
%
\vspace*{5pt}\begin{verbatim}
msmmAltMoments <- function(theta,x){
# extract variables from x
Y <- x[,"y"]
X <- x[,"x"]
Z1 <- x[,"z1"]
Z2 <- x[,"z2"]
# moments
m1 <- (Y*exp(-theta[1] - X*theta[2]) - 1)
m2 <- (Y*exp(-theta[1] - X*theta[2]) - 1)*Z1
m3 <- (Y*exp(-theta[1] - X*theta[2]) - 1)*Z2
return(cbind(m1,m2,m3))
}

msmm2 <- gmm(msmmAltMoments, x=data,
t0=c(0,0), vcov="iid")
print(exp(cbind(coef(msmm2),
confint(msmm2)))) # exponentiate estimates
print(specTest(msmm2))
\end{verbatim}
%
%EndExpansion
}

\subsubsection*{Logistic SMM}

In estimation of the logistic SMM, especially with the joint moments,
it is
important to check that convergence has been reached, either by inspecting
the model summary or checking that the model \texttt{algoInfo\$convergence}
attribute is equal to 0. If convergence has not been reached, a higher
iteration limit (say, 5000) can be specified in \texttt{gmm()} through the
option \texttt{control=list(maxit=}\texttt{5000)}. Note in R $\operatorname {qlogis}(p)=\log
(p/(1-p))$ and $\operatorname{plogis}(x)=\expit (x)=e^{x}/(1+e^{x})$.

First we fit the association model using the \texttt{glm()} function
to fit
the logistic regression. Again we collect the parameter estimates and
predicted values. We then fit the causal model using the function
\texttt{cmMoments()} to return its moment conditions. In this function \texttt{theta[1]} denotes $E ( Y_{0} ) $ and \texttt{theta[2]}
denotes the
log causal odds ratio.

In the joint estimation the function \texttt{lsmmMom\-ents()} returns the
moment conditions. In this function \texttt{theta[1:6]} are the coefficients
in the association model, \texttt{theta[7]} denotes $E (
Y_{0} ) $
and \texttt{theta[8]} denotes the log causal odds ratio.

{\footnotesize
%TCIMACRO{%
%# association model
%am <- glm(y ~ x + z1 + z2 + x*z1 + x*z2, as.data.frame(data),
%fam=binomial)
%print(summary(am))
%amfit <- coef(am)
%xblog <- qlogis(fitted.values(am))
%
%# causal model with incorrect SEs
%cmMoments <- function(theta,x){
%# extract variables from x
%X <- x[,"x"]
%Z1 <- x[,"z1"]
%Z2 <- x[,"z2"]
%# moments
%c1 <- (plogis(xblog - theta[2]*X) - theta[1])
%c2 <- (plogis(xblog - theta[2]*X) - theta[1])*Z1
%c3 <- (plogis(xblog - theta[2]*X) - theta[1])*Z2
%return(cbind(c1,c2,c3))
%}
%
%cm <- gmm(cmMoments, x=data, t0=c(0,0), vcov="iid")
%cmfit <- coef(cm)
%
%lsmmMoments <- function(theta,x){
%# extract variables from x
%Y <- x[,"y"]
%X <- x[,"x"]
%Z1 <- x[,"z1"]
%Z2 <- x[,"z2"]
%XZ1 <- X*Z1
%XZ2 <- X*Z2
%# association model moments
%xb <- theta[1] + theta[2]*X + theta[3]*Z1 + theta[4]*Z2 + theta[5]*XZ1
%+ theta[6]*XZ2
%a1 <- (Y - plogis(xb))
%a2 <- (Y - plogis(xb))*X
%a3 <- (Y - plogis(xb))*Z1
%a4 <- (Y - plogis(xb))*Z2
%a5 <- (Y - plogis(xb))*XZ1
%a6 <- (Y - plogis(xb))*XZ2
%# causal model moments
%c1 <- (plogis(xb - theta[8]*X) - theta[7])
%c2 <- (plogis(xb - theta[8]*X) - theta[7])*Z1
%c3 <- (plogis(xb - theta[8]*X) - theta[7])*Z2
%return(cbind(a1,a2,a3,a4,a5,a6,c1,c2,c3))
%}
%
%lsmm <- gmm(lsmmMoments, x=data, t0=c(amfit,cmfit), vcov="iid")
%print(summary(lsmm))
%print(cbind(coef(lsmm), confint(lsmm))[8]) # E[Y(0)]
%print(exp(cbind(coef(lsmm), confint(lsmm))[-7,])) # exponentiate other
%estimates
%print(specTest(lsmm))
%BeginExpansion
%
\vspace*{5pt}\begin{verbatim}
# association model
am <- glm(y ~ x + z1 + z2 + x*z1 + x*z2,
as.data.frame(data), fam=binomial)
print(summary(am))
amfit <- coef(am)
xblog <- qlogis(fitted.values(am))

# causal model with incorrect SEs
cmMoments <- function(theta,x){
# extract variables from x
X <- x[,"x"]
Z1 <- x[,"z1"]
Z2 <- x[,"z2"]
# moments
c1 <- (plogis(xblog - theta[2]*X)
- theta[1])
c2 <- (plogis(xblog - theta[2]*X)
- theta[1])*Z1
c3 <- (plogis(xblog - theta[2]*X)
- theta[1])*Z2
return(cbind(c1,c2,c3))
}

cm <- gmm(cmMoments, x=data, t0=c(0,0),
vcov="iid")
cmfit <- coef(cm)

lsmmMoments <- function(theta,x){
# extract variables from x
Y <- x[,"y"]
X <- x[,"x"]
Z1 <- x[,"z1"]
Z2 <- x[,"z2"]
XZ1 <- X*Z1
XZ2 <- X*Z2
# association model moments
xb <- theta[1] + theta[2]*X + theta[3]*Z1
+ theta[4]*Z2 + theta[5]*XZ1
+ theta[6]*XZ2
a1 <- (Y - plogis(xb))
a2 <- (Y - plogis(xb))*X
a3 <- (Y - plogis(xb))*Z1
a4 <- (Y - plogis(xb))*Z2
a5 <- (Y - plogis(xb))*XZ1
a6 <- (Y - plogis(xb))*XZ2
# causal model moments
c1 <- (plogis(xb - theta[8]*X)
- theta[7])
c2 <- (plogis(xb - theta[8]*X)
- theta[7])*Z1
c3 <- (plogis(xb - theta[8]*X)
- theta[7])*Z2
return(cbind(a1,a2,a3,a4,a5,a6,c1,c2,c3))
}

lsmm <- gmm(lsmmMoments, x=data,
t0=c(amfit,cmfit),
vcov="iid")
print(summary(lsmm))
print(cbind(coef(lsmm), confint(lsmm))[8])
# E[Y(0)]
print(exp(cbind(coef(lsmm),
confint(lsmm))[-7,])) # exponentiate other
estimates
print(specTest(lsmm))
\end{verbatim}
%
%EndExpansion
}
\end{appendix}\iffalse\fi
\section*{Acknowledgments}
The authors would like to thank B{\o}rge Nordestgaard for access
to the Copenhagen General Population Study data. We also thank two anonymous
referees and the Editor for very useful comments which improved the
manuscript, and George Davey Smith, Nicholas Timpson, Vanessa Didelez, Roger
Harbord, Nuala Sheehan and conference participants in London, Lund, Malaga,
Manchester and Mannheim for helpful comments.

Research supported in part by UK Economic \& Social Research Council grants
RES-060-23-0011 and RES-576-25-0035, UK Medical Research Council grants
G0601625 and G0600705, and European Research Council grant 269874-DEVHEALTH.

%suskaldyti doi

% imsref loaded by imikolaityte, 2014-12-09 11:26:30
%

\end{document}